\documentclass[prd,aps,amsfonts,showpacs,superscriptaddress,nofootinbib,longbibliography,notitlepage,twocolumn]{revtex4-1}

\usepackage{graphicx}
\usepackage[usenames,dvipsnames]{xcolor}
\usepackage{rotating}
\usepackage[tight]{subfigure}
\usepackage[normalem]{ulem}
\usepackage{amsmath,amssymb,graphics,amsthm}
\usepackage{appendix}

\usepackage[colorlinks=true, urlcolor=violet, linkcolor=blue, citecolor=red, hyperindex=true, linktocpage=true]{hyperref}
\usepackage[capitalise,compress]{cleveref}
\usepackage{soul}

\makeatletter
\renewcommand{\p@subsection}{}
\renewcommand{\p@subsubsection}{}
\makeatother

\allowdisplaybreaks

\usepackage{xcolor}

\usepackage{mathtools}

\newcommand\p{\ensuremath{\partial}}
\newcommand{\be}{\begin{equation}}
\newcommand{\ee}{\end{equation}}

\begin{document}
\title{Breakdown of hydrodynamics below four dimensions in a fracton fluid}

\author{Paolo Glorioso}
\email{paolog@stanford.edu}
\affiliation{Department of Physics, Stanford University, Stanford, CA 94305, USA}

\author{Jinkang Guo}
\affiliation{Department of Physics and Center for Theory of Quantum Matter, University of Colorado, Boulder, CO 80309, USA}

\author{Joaquin F. Rodriguez-Nieva}
\affiliation{Department of Physics, Stanford University, Stanford, CA 94305, USA}

\author{Andrew Lucas}
\email{andrew.j.lucas@colorado.edu}
\affiliation{Department of Physics and Center for Theory of Quantum Matter, University of Colorado, Boulder, CO 80309, USA}

\begin{abstract}
We present the nonlinear fluctuating hydrodynamics which governs  the late time dynamics of a chaotic many-body system with simultaneous charge/mass, dipole/center of mass, and momentum conservation. This hydrodynamic effective theory is unstable below four spatial dimensions: dipole-conserving fluids at rest become unstable to fluctuations, and are governed not by hydrodynamics, but by a fractonic generalization of the Kardar-Parisi-Zhang universality class.  We numerically simulate many-body classical dynamics in one-dimensional models with dipole and momentum conservation, and find evidence for a breakdown of hydrodynamics, along with a new universality class of undriven yet non-equilbrium dynamics. 
\end{abstract}

\date{\today}

\maketitle

\section{Introduction} One of the oldest and most applicable theories in physics is hydrodynamics.  While hydrodynamics was first understood as a phenomenological set of equations that govern liquids and gases \cite{landau}, over the past century we have instead recognized that hydrodynamics is best understood as the universal effective field theory that governs thermalization in a chaotic many-body system \cite{Crossley:2015evo,Haehl:2015foa,Jensen:2017kzi}.  Due to this universality, the same theories of hydrodynamics can describe diverse phases of classical or quantum matter, including ultracold atoms \cite{Cao_2010}, quark-gluon plasma \cite{Shuryak_2009}, and electrons and phonons in high-purity solids \cite{Crossno_2016,Bandurin_2016,de_Jong_1995,Krishna_Kumar_2017}.

Novel  phases of matter arise when the microscopic degrees of freedom are \emph{fractons} -- excitations which are individually immobile, and can only move in tandem \cite{chamon2005quantum,vijay2015new,vijay2016fracton, pretko2017subdimensional,pretko2017emergent,pretko2017generalized,slagle2017fracton,slagle2018symmetric, schmitz2018recoverable,ma2017fracton,ma2018topological,shirley2019foliated,slagle2017fracton,Pretko2018,radzihovsky2020fractons,Seiberg_2020,Gorantla_2020,rudelius2020fractons,doshi}.   As a simple example, we can consider a phase of matter in which the global charge (or mass) is conserved, together with the global dipole moment (or momentum). In this case, a single particle cannot move without violating the dipole conservation law.  If such a phase of matter, realized on a lattice, can thermalize \cite{pai2019localization,2020PollmannFragmentation,shattering}, it is described by a novel hydrodynamics in which Fick's law of charge diffusion is replaced by slower subdiffusion \cite{fractonhydro,knap2020,zhang2020,morningstar,iaconis2021,Ganesan:2020wvm}.  The emergence of subdiffusion is not special to peculiar microscopic details of particular lattice models; it is guaranteed by the symmetries of the dynamics.  This robustness of hydrodynamics to microscopic peculiarities makes it an experimentally ideal probe for constrained dynamics \cite{Guardado_Sanchez_2020}. 

Here, we study such a dipole-conserving theory which is also translation-invariant.  In this case, charge, dipole and momentum are all conserved quantities.   We show that these fluids exhibit a highly unusual hydrodynamics, with magnon-like propagations with subdiffusive decay rates.  More importantly, below four spatial dimensions, these fluids are violently unstable to thermal fluctuations.  Hydrodynamics thus will not exist in any experimentally-realizable spatial dimension; rather, it will be replaced by a fractonic generalization of the Kardar-Parisi-Zhang (KPZ) universality class \cite{kpz}.

In one spatial dimension, the KPZ fixed point is remarkably generic: it arises in the study of growing surfaces \cite{kpz}, quantum Hall edge states \cite{gloriosoprl}, and -- most importantly for us -- is the endpoint of a fundamental instability of the Navier-Stokes equations in one spatial dimension \cite{Spohn_2014}.  Among the many non-equilibrium universality classes that have been discovered in statistical mechanics, ranging from flocking \cite{flocking} in active matter, to driven-dissipative condensates \cite{PhysRevResearch.2.033018}, to fluctuating smectic liquid crystals \cite{toner}, the KPZ fixed point is unique in that it describes the instability of an ordinary undriven fluid at rest, without any spontaneous symmetry breaking.  By studying the curious hydrodynamics arising in matter with fractonic conservation laws, we have  discovered that such instabilies of fluids in equilibrium can exist in three dimensions as well.  

Just as hydrodynamics is robust against microscopic details, so too is the universality class of non-equilibrium dynamics that emerges out of a hydrodynamic instability.  The universality class of a dipole-conserving fluid can be realized in any medium with exact or emergent dipole and momentum conservation.   While we are not aware of a current experimental platform exhibiting these conservation laws, we outline possible strategies to building them in what follows.  Independently of experiment, it is possible to realize these conservation laws in classical or quantum dynamics which can be simulated numerically.   We have numerically simulated one-dimensional chaotic classical dynamics with dipole and momentum conservation, and find evidence for the advertised fractonic generalization of the KPZ fixed point.  Our work thus establishes an unexpected and profound connection between non-equilibrium statistical mechanics and unconventional fracton phases of matter.

\section{Microscopic Models}Given the seemingly abstract nature of how fluids might simultaneously have both dipole and momentum conservation, before diving into the theoretical framework of fluid dynamics, it is instructive to first describe a microscopic model which would lie in such a universality class.  For simplicity in the discussion which follows, we focus on one dimensional systems.   Let us begin by considering $N$ particles with momenta $p_i$ and displacements $x_i$ ($i=1,\ldots,N$), coupled together by the following Hamiltonian: \begin{equation}
    H = \sum_{n=1}^{N-1} \frac{(p_n-p_{n+1})^2}{2} + V(x_n - x_{n+1}). \label{eq:Hdip}
\end{equation}
Here $V(x) = V_2 x^2 + V_3 x^3 + \cdots$ is a generic polynomial.   This is qualitatively similar to a simple model of one-dimensional solids with anharmonicity, except for the kinetic energy, which depends on only the difference of momenta.  Somewhat similar models have arisen in the ``dipole fermion" picture of fractional quantum Hall states \cite{simon,fliss2021entanglement}.

This curious kinetic energy is all that we need to have an emergent dipole conservation.  Using the Poisson brackets $\lbrace x_n, p_m\rbrace = \delta_{nm}$, we find that the dipole moment $D$, ``charge" $Q$ and momentum $P$, given by \begin{equation}
    Q = \sum_{n=1}^N 1,  \;\;\;
    D = \sum_{n=1}^N x_i,  \;\;\;
    P = \sum_{n=1}^N p_i,
\end{equation}
obey the classical multipole algebra \cite{gromov2018towards} \begin{equation} \label{eq:multipole}
    \lbrace Q,D\rbrace = \lbrace Q, P\rbrace = 0, \;\;\; \lbrace D,P\rbrace = Q.
\end{equation}
Since $Q,D$ and $P$ commute with the Hamiltonian,
%$\lbrace Q,H\rbrace = \lbrace D,H\rbrace = \lbrace P,H\rbrace = 0$, 
we have conservation of charge, dipole and momentum all at once, in a spatially local theory.  Note that energy is also conserved if the dynamics is Hamitonian.  But it is straightforward to modify the dynamics to no longer conserve energy.

If we assume the dynamics is close to equilibrium (i.e. displacements are small), then we can safely set (without loss of generality) $V(x) = \frac{1}{2}x^2$.  A simple calculation reveals that the normal modes of this quadratic and integrable system are of the form $(x_n,p_n) \propto \mathrm{e}^{\mathrm{i}kn - \mathrm{i}\omega_k t}$, and that when $k \ll 1$, $\omega_k \approx \pm k^2$.  As promised in the introduction, we have uncovered a magnon-like dissipationless dispersion relation, which we will show later is universal and follows entirely from simultaneous dipole and momentum conservation, even when we include higher order terms (which destroy integrability) in $V(x)$.

As we have found a magnon-like dispersion relation, it is tempting to push the analogy between dipole conserving fluids, and isotropic ferromagnets, a little further.   Consider the isotropic Heisenberg ferromagnet \begin{equation}
    H = -\sum_{n=1}^{N-1}\mathbf{S}_n \cdot \mathbf{S}_{n+1},
\end{equation}
where $\mathbf{S}_n = (S^x_n, S^y_n, S^z_n)$ and $\lbrace S^i_n,S^j_m\rbrace = \epsilon^{ijk}S^k_n \delta_{nm}$.  Imposing the constraint $S^x_n S^x_n+S^y_n S^y_n+S^z_n S^z_n=1$, and perturbing around the minimal energy configuration $S^z_n=1$, we observe that if we identify charge $Q$, dipole $D$ and momentum $P$ with the total $x$, $y$ and $z$ components of spin, the spin algebra is equivalent to (\ref{eq:multipole}).  Moreover, Taylor expanding $H$ to leading order in small $S^{x,y}_n$, we observe that up to a global constant, $H$ is approximately given by (\ref{eq:Hdip}) with $V(x)=\frac{1}{2}x^2$.  

Remarkably, we see that the isotropic ferromagnet has an \emph{approximate} dipole and momentum conservation close to equilibrium.  However, nonlinearities in the ferromagnet do not preserve $S^z_n=1$, and so the nonlinear theories differ.  Despite this nonlinear discrepancy, we note that nonlinearities are known to be strongly relevant below six dimensions in the Heisenberg ferromagnet \cite{hohenberg}.    In the exact dipole-conserving theory, we will show that the upper critical dimension is four.  And while the Mermin-Wagner theorem destroys order in the ferromagnet in one dimension, there is no destruction of dipole conservation in the model (\ref{eq:Hdip}).  It would be interesting to investigate whether a ferromagnet can be modified by realizable interactions to better stabilize the dipole-conserving hydrodynamics.

\section{Hydrodynamics}We now use canonical arguments, based on the second law of thermodynamics, to derive the hydrodynamics of conserved momentum, charge, and dipole. The fundamental assumption of hydrodynamics is that the late time physics is governed locally by the independent quantities of the system, which we write as
\be P^i = \int \mathrm{d}^d x\, \pi^i,\quad Q=\int \mathrm{d}^d x\, n \ \ee
where $\pi^i$ and $n$ are the momentum and charge density, respectively.  $ij$ indices in what follows run over spatial dimensions $i=1,\ldots d$, and repeated indices are  summed over. The dynamics of the densities $n$ and $\pi^i$ is given by the local conservation laws:
\be \label{eom1}\p_t \pi^i + \p_j T^{ji}=0,\qquad \p_t n+\p_i J^i=0\ ,\ee
where $T^{ij}$ and $J^i$ are stress and charge flux, and are assumed to be local expressions of $\pi^i,n$. Crucially, we also need to demand
\be\label{dipfl} J^i = \p_j J^{ji}\ ,\ee
which comes from dipole conservation:
\be \label{dipfl1}\p_t \int d^d x x^i n= -\int \mathrm{d}^d x x^i\p_j J^j=\int \mathrm{d}^d x J^i\ ,\ee
where the right-hand side vanishes only if $J^i$ satisfies (\ref{dipfl}). We will also demand that $J^{ij}$ be local in the densities. 
%We will mention a symmetry-based argument below that leads to (\ref{dipfl}) without assuming (\ref{dipfl1}). 

We have not included the dipole density as a separate degree of freedom. Indeed, let us decompose the dipole charge as: $D^i=D_0^i+S^i$, where $D_0^i=\int \mathrm{d}^d x\, x^i n$ is the ``orbital'' component and $S^i$ a remainder, corresponing to a density of microscopic dipoles.  In general, we only expect the sum $D^i$ to be conserved, and not each component separately.  Therefore, the $S^i$ charge will typically relax into the local density $x^in$, and dipole density is not a separate hydrodynamic degree of freedom \cite{Glorioso_2021}.  This is analogous to the reason why a fluid with angular momentum conservation does not have a new hydrodynamic mode associated to angular momentum density.

Upon specifying the explicit dependence of $T^{ij}$ and $J^i$ on $\pi^i$ and $n$, Eqs. (\ref{eom1}),(\ref{dipfl}) will completely specify the time evolution of $\pi^i$ and $n$. To find such explicit dependence, we shall write down the most general expressions of $T^{ij}$ and $J^i$ in terms of $\pi^i$ and $n$ following a derivative expansion, and then impose that the dynamics be consistent with the local second law of thermodynamics. This amounts at finding a vector $J_S^i$ such that 
\be \label{entp}\p_t s+\p_i J_S^i\geq 0\ee
when evaluated on solutions to hydrodynamics, where $s$ is the thermodynamic entropy density. This basic constraint will uniquely determine the concrete expressions of $T^{ij},J^i$ in terms of $\pi^i,n$, order by order in derivatives, up to phenomenological coefficients that are determined by the specific underlying system. 

The thermodynamics and hydrodynamics of dipole-conserving systems are special: contrary to all cases known to us, the homogeneous part of momentum density decouples from the dynamics. Indeed, we begin by first assuming that the entropy density is a function of momentum and charge densities $s=s(\pi^i,n)$, as in conventional thermodynamics.  We now show that this breaks (\ref{dipfl}). Recall the thermodynamic relation \begin{equation}
    T\mathrm{d}s= -V^i \mathrm{d}\pi^i-\mu \mathrm{d}n,
\end{equation} where $V^i$ and $\mu$ are the velocity and chemical potential of the system. We recall that we are assuming absence of energy conservation, so we will take $T$ to be a constant set by noise in this discussion, and will study hydrodynamics with energy conservation in a technical companion paper \cite{future}.
%\PG{(say more about this)}, moreover we take the temperature $T$ to be a constant.\footnote{In the examples of the previous section, the noisy model has temperature set by the variance of the noise. In the tilted model the energy density becomes a relaxed mode due to Joule heating, thus making temperature effectively constant.} [ANDY: concerned about this footnote: in tilted models I think $T=\infty$, arguably.  my personal view is that it's best to just say that we assume noise fixes constant $T$, and talk about tilting later] [Paolo: the phase space of our tilted model is unbounded, wouldn't $T=\infty$ be singular?] [ANDY: don't think so once $D$ is fixed] 
Combining this thermodynamic relation with the fact that, in non-dissipative hydrodynamics, entropy is locally conserved, we arrive at
\be \label{2ndla0}T(\p_ts+\p_i J_S^i) =-V^i(\p_t\pi^i+\p_j T^{ji}) - \mu(\p_tn+\p_i J^i). \ee
The most general expressions for the currents are $J_S^i = s_1 V^i$, $T^{ij} = p \delta^{ij} + h_1 \pi^i V^j$, $J^i = h_2 V^i$, where $s_1,p,h_1,h_2$ are functions of $\pi^i$ and $n$. Plugging these expressions in (\ref{2ndla0}) gives $s_1=s$, $p=Ts+\mu n+ V^i \pi^i$, $h_1 V^i =\pi^i$ and $h_2=n$, which are just the standard constitutive relations of the hydrodynamics of a charged fluid: indeed, we have not used anywhere the fact that we are dealing with a dipole-conserving fluid. In particular, these results together with (\ref{dipfl}) lead to the relation 
\be\label{nV1} n V^i = \p_j J^{ji},\ee
which would naively imply that $J^{ij}$ is non-local in the hydrodynamic variables, thus violating our basic assumptions. %Eq. (\ref{nV1}) is Bloch's theorem and is a basic consequence of the global U(1) symmetry of the underlying system associated to the conservation of $n$.

The only way for (\ref{nV1}) to be consistent with locality, therefore, is to demand that the velocity $V^i$ is itself a total derivative (divided by $n$). Since $V^i$ is defined as the chemical potential of $\pi^i$, such requirement is only possible if the entropy density has the following dependence on $\pi^i$:
\be s=s(\p_i v_j,n),\qquad v_i\equiv\frac{\pi^i}{n}\label{td2}\ .\ee
Again demanding (\ref{2ndla0}), we find 
\begin{subequations}\begin{align}
     T^{ij} &= p \delta^{ij} + V^i \pi^j - \psi_{ik}\partial_j v_k, \label{const1} \\
     J^{ij} &= \psi_{ij},\label{const1a}
    \end{align}\end{subequations}  
where the velocity and the thermodynamic pressure are
\begin{equation}
    V^i=\frac 1n \p_j \psi_{ji},\qquad p=Ts-nT\frac{\partial s}{\partial n} \label{const2},
\end{equation}
and we have defined the quantity
\begin{equation}
    \psi_{ij}\equiv\left.T\frac{\p s}{\p(\p_i v_j)}\right|_n\,.\label{const4}
\end{equation}
Unlike ordinary fluids, the velocity is a higher-derivative expression of momentum density. This is the only way to reconcile (\ref{nV1}) with locality. In a rotationally invariant theory, we find that $T_{ij}$ is symmetric up to total derivatives, consistent with conservation of angular momentum.  An explicit derivation of these facts is provided in Appendix \ref{app:zero}.

Note that the entropy density $s$ as well as equations of motion are invariant under the shift 
\be\label{dipsh} \pi^i\to \pi^i + n c^i,\qquad T^{ij}\to T^{ij} + J^j c^i\ ,\ee where $c^i$ is a constant vector. This invariance is a manifestation of the dipole algebra (\ref{eq:multipole}). Indeed, (\ref{eq:multipole}) implies, using locality: $\lbrace D^i,\pi^j \rbrace= n\delta^{ij}$, which is equivalent to the symmetry (\ref{dipsh}). In fact, using (\ref{dipsh}) as the only input, one immediately infers  (\ref{td2}), which in turn imply (\ref{const1})-(\ref{const4}) and in particular (\ref{dipfl}). Such symmetry-based approach confirms that the hydrodynamics (\ref{const1})-(\ref{const4}) is valid for arbitrary strongly-coupled systems and thus universal.

It is straightforward to derive first order dissipative corrections to hydrodynamics.  We do this calculation in Appendix \ref{app:const}, and now summarize the results.  We find that $J^{ij} = J^{ij}_{(0)} + J^{ij}_{(1)}$ and $T^{ij} = T^{ij}_{(0)} + T^{ij}_{(1)}$, where $J^{ij}_{(0)}$ and $T^{ij}_{(0)}$ correspond to the ideal hydrodynamic results derived above, and \begin{subequations}\label{con1}\begin{align} -T^{ij}_{(1)}&=\eta^{ijkl}\p_k V_l+\alpha^{ijkl}\p_k\p_l \mu\\
J^{ij}_{(1)}&=\kappa^{ijkl}\p_k V_l+C^{ijkl}\p_k\p_l \mu.\end{align}\end{subequations}
The tensor structures are detailed in Appendix \ref{app:const}, and are similar to shear and bulk viscosities of an isotropic fluid.

To complete our hydrodynamic description, we finally add the effect of thermal fluctuations. Generalizing the standard fluctuation-dissipation theorem \cite{landau}, we add noise to the currents: $T^{ij}\to T^{ij}+\tau^{ij}$ and $J^{ij}\to J^{ij}+\xi^{ij}$, where the variance is determined by the dissipative coefficients of (\ref{con1}):
\begin{subequations}\label{noise}\begin{align} \langle \tau^{ij} \tau^{kl}\rangle&=2T \eta^{ijkl}\delta(t)\delta^{(d)}(x)\\
\langle \xi^{ij} \xi^{kl}\rangle&=2T C^{ijkl}\delta(t)\delta^{(d)}(x)\\
\langle \tau^{ij} \xi^{kl}\rangle&=T(\alpha^{ijkl}+\kappa^{klij})\delta(t)\delta^{(d)}(x)
\end{align}\end{subequations}
In Appendix \ref{app:linearized}, we derive the propagating hydrodynamic modes of this theory: namely, we look for solutions to the hydrodynamic equations in which $n,v_i \propto \mathrm{e}^{\mathrm{i}kx-\mathrm{i}\omega t}$.  We find a magnon-like ``sound'' mode with dispersion relation  \cite{forster,hohenberg} $\omega = \pm ck^2 - \mathrm{i}\gamma k^4$, and $d-1$ subdiffusive modes for transverse momentum with dispersion relation $\omega = -\mathrm{i}\gamma^\prime k^4$.  Explicit expressions for $c,\gamma,\gamma^\prime$ are not illuminating and are provided in the appendix.  Note that the qualitative structure of these quasinormal modes matches those of an ordinary fluid, except that each power of wave number $k$ is doubled. 

\section{Instability of Hydrodynamics}
In fact, the true dispersion relations differ from those we found from linear response above. Relevant nonlinearities couple to thermal fluctuations and lead to anomalous scaling,  severely affecting the long-time behavior of general dipole-conserving hydrodynamics. For simplicity, we shall present the explicit nonlinearities only in one dimension; the higher dimensional counterpart is qualitatively similar and can be found in Appendix \ref{app:expl}. 
%Consider the entropy density up to cubic order in $\delta n$:
%\be \label{ent2}s=-\frac 12 a^{ijkl}\p_i v_j \p_k v_l-\frac{\mu_0}T\delta n-\frac 1{2\chi T}\delta n^2-\frac {\lambda}{6 T}\delta n^3,\ee
%where we assume that there are higher order terms in $\delta n$ that guarantee thermodynamic stability. 
We consider perturbations of an equilibrium fluid at rest --  $v_x=0$ and $n=n_0$ ($\delta n = n -n_0$) -- and find
\begin{subequations}\begin{align} &\p_t v_x+\frac 1\chi \p_x\delta n+\lambda\delta n\p_x\delta n+\lambda^\prime\delta n^2 \p_x\delta n\nonumber\\
&+\frac{aT}{n_0^2}\Gamma\p_x^4v_x-\frac A{n_0\chi}\p_x^3 \delta n+\frac 1{n_0}\p_x\tau^{xx}+\cdots=0,\\
&\p_t \delta n-A\p_x^3 v_x+\frac C\chi\p_x^4 \delta n+\p_x^2\xi^{xx}+\cdots=0.
\end{align}\end{subequations}
$\cdots$ denote higher derivative/nonlinear terms which are not important for what follows. 
%[Paolo: since below we discuss scaling in generic dimension, it would be nice if we can still quote the equations in general dimension. I wrote them below using vector notation. Can we further simplify them, perhaps by writing them in a schematic way??]
%\begin{subequations}\begin{align} &\p_t v+\frac 1\chi \nabla\delta n+\lambda\delta n\nabla\delta n+\frac {T}{n_0}(a-a_2-a_3)\nabla(\nabla\cdot v)\times\nabla\times v\nonumber\\
%&+\frac{T}{n_0}(a_2+a_3)\nabla^2 v\times \nabla\times v+\frac {aT}{n_0^2}\Gamma_1\nabla^4v\nonumber\\
%&+\frac {aT}{n_0^2}\Gamma_2 \nabla^2\nabla( \nabla\cdot v)
%-\frac A{n_0\chi} \nabla\p^2 \delta n+\frac 1{n_0}\zeta_\pi=0\\
%&\p_t \delta n-aT \p^2\nabla\cdot v+\frac C\chi \nabla^4 \delta n-\frac {aT}{n_0}A\nabla^4\nabla\cdot v+\zeta_n=0
%\end{align}\end{subequations}
%[Paolo: setting $a_2,a_3,\Gamma_2,A=0$, the equations still keep the essential features:]
%\begin{align} \p_t v=&-\frac 1\chi \nabla\delta n-\lambda\delta n\nabla\delta n\nonumber\\&-\frac {aT}{n_0}\nabla(\nabla\cdot v)\times\nabla\times v-\frac {aT}{n_0^2}\Gamma_1\nabla^4v-\frac 1{n_0}\zeta_\pi\nonumber\\
%\p_t \delta n=&aT \p^2\nabla\cdot v-\frac C\chi \nabla^4 \delta n-\zeta_n
%\end{align}
We included stochastic fluctuations in the equations.  The values of constants $\lambda$, $\lambda^\prime$, $\chi$, $a$, $A$ and $C$ do not depend on $v_x$ or $\delta n$.

Neglecting fluctuations, the dissipative scaling is $\omega\sim k^4$. From (\ref{noise}), the noise scales as $\tau,\eta\sim k^{\frac{d+4}2}$. Starting from the linearized theory and assuming that $a$, $A$ etc. are scale ($k$) independent, we find $\pi^i\sim k^{\frac{d-2}2}$ and $n\sim k^{\frac d2}$.  As per the usual renormalization group analysis, the coefficient of $\lambda$ must scale as $k^{\frac{d-4}2}$,  making it \emph{relevant} when $d<4$; $\lambda^\prime$ scales as $k^{d-2}$ and is relevant when $d<2$.  As a consequence, we can anticipate anomalous dissipative scaling: the magnon-like sound mode will have dispersion relation $\omega\sim k^2-ik^z$, with $z<4$.  We crudely estimate $z$ by assuming that, even after fluctuations are accounted for, the scaling of the densities does not renormalize. Assuming $\lambda\ne 0$ and balancing the time derivative with nonlinearities $\p_t \pi^i\sim \nabla (\delta n)^2,\nabla(\nabla\pi)^2$ leads to $k^{z+\frac{d-2}2}\sim k^{d+1}$, or $z\sim d/2+2$. In one dimension, this gives $z\sim 2.5$. 
%In one dimension nonlinearities proportional to $a$ vanish and thus, if the underlying microscopic system has particle-hole symmetry $n\to -n$, the leading nonlinearity comes from the pressure term in (\ref{const1}) and is proportional to $\delta n^3$. This nonlinearity is still relevant but leads to a smaller deviation from $z=4$, consistently with Fig. \ref{fig:z4}. 

\section{Numerical Simulations} 
Having predicted both the exotic dissipative hydrodynamics of a dipole-conserving theory, together with its breakdown due to thermal fluctuations, we now describe two models which we have used to numerically test our predictions.

\textbf{Model A:} Our first model starts with Hamiltonian (\ref{eq:Hdip}), with
\begin{align}
V(x) &= \frac{1}{2}x^2 + k_3 x^3 + k_4 x^4. \label{eq:Vx}
\end{align}
Note that Model A has energy conservation, and strictly speaking our hydrodynamic derivation above does not.  Energy conservation changes the universality class and critical exponents of hydrodynamics \cite{future}, and so strictly speaking, model A does not lie in the universality class predicted above.  In Appendix \ref{app:modelA}, we study a stochastic version of model A with noise and dissipation, which is not energy conserving and is predicted to lie in our universality class.  Further simulations, and discussion of the role of energy conservation, are also described there.

\textbf{Model B:} Our second model corresponds to  \begin{equation}
    H = \sum_{i=1}^N \left(-\cos p_i - F x_i\right) + V(x),
\end{equation}
with $V(x)$ given in (\ref{eq:Vx}).  Note that this model does \emph{not} have explicit dipole conservation, nor momentum conservation.  However, analogous to the emergence of dipole conservation out of energy conservation in systems placed in strong tilt fields \cite{fractonhydro,zhang2020universal,Guardado_Sanchez_2020}, we predict the emergence of dipole \emph{and} momentum conserving hydrodynamics in this model.   Indeed, in Appendix \ref{app:modelB}, we explicitly show that the linearized equations in model B exhibit fast Bloch oscillations superimposed on top of magnon-like hydrodynamic modes.  Model B is inspired by one possible experimental realization of dipole-conserving hydrodynamics, in which ultracold fermionic atoms are placed in a tilted optical lattice \cite{Guardado_Sanchez_2020}.  The $\cos p_i$ kinetic energy arises from the finite bandwidth of a lattice model, and the $Fx_i$ force field comes from the tilt.  In order for this realization to be appropriate, it is important for umklapp scattering to be suppressed and for momentum to be approximately conserved.

\begin{figure}
\includegraphics[scale = 1.0]{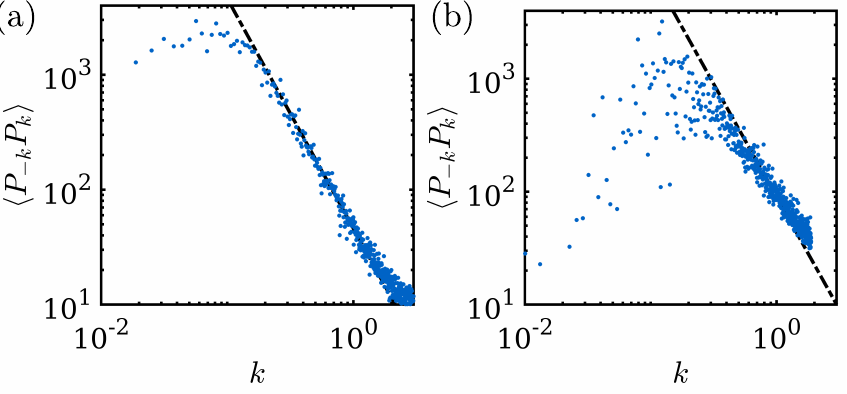}
\caption{Distribution function of $\langle p_{-k}(t)p_k(t)\rangle$ correlations at late times for (a) model A and (b) model B. The dashed line is the theoretical $k^{-2}$ prediction.}
\label{fig:k2}
\end{figure}

We now present large scale simulations for each model.  A ``thermodynamic" check for dipole-conserving hydrodynamics is to look at the equilibrium fluctuations of the momentum density, which are proportional to the momentum susceptibility: using (\ref{nV1}) and (\ref{const2}), \begin{equation}
    \chi_{PP} = \frac{\pi_i}{V_i} \propto k^{-2}.
\end{equation}
As $k\rightarrow 0$, we predict a clear divergence in the equal time correlation functions $\langle p_k(t)p_{-k}(t)\rangle$, with $\langle\cdots\rangle$ an average over times and/or initial conditions; here $p_k$ denotes the discrete Fourier transform of $p_i$.  Figure \ref{fig:k2} demonstrates this divergence is present in both models.

%\PG{(remove the following:)} Next, we study the relaxation time scale of each model.   Initializing each model with random initial conditions, 

\begin{figure}
\includegraphics[scale = 1.0]{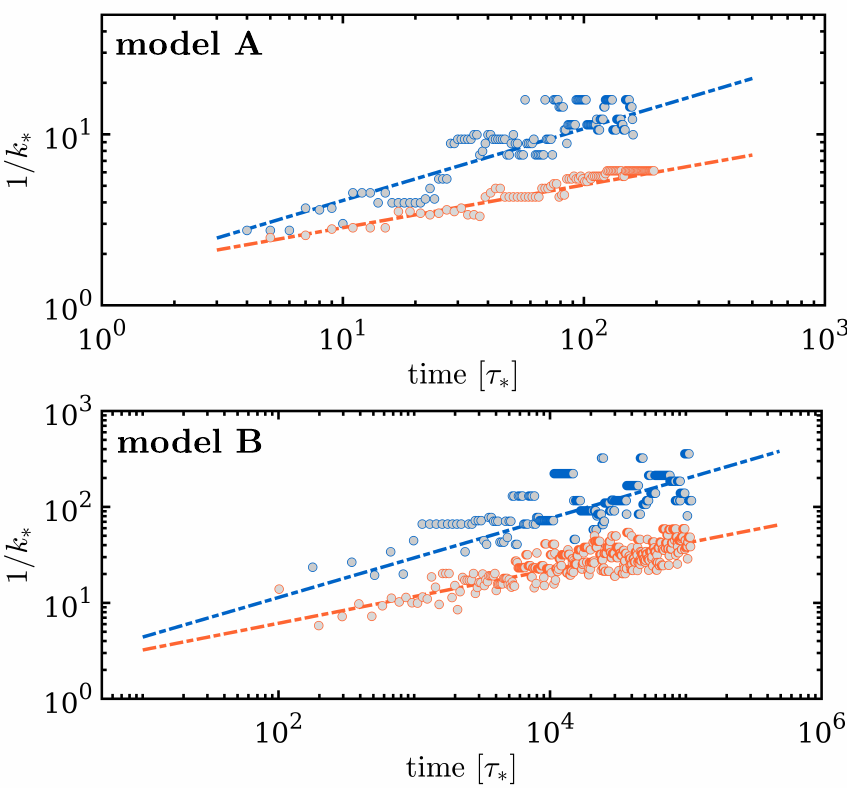}

\includegraphics[scale = 1.0]{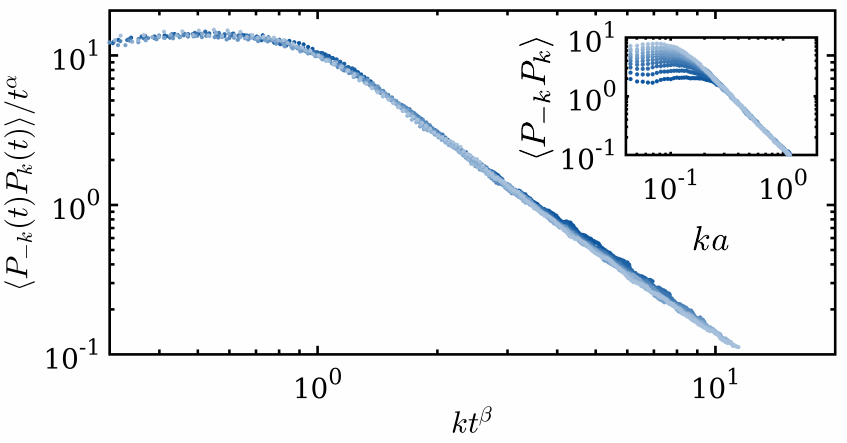}
\caption{Temporal dependence of $1/k_*$ showing anomalous scaling in model A (\textbf{top}) and model B (\textbf{middle}). Shown with blue (orange) is the case when $k_3\ne 0$ ($k_3=0$). We estimate $1/z = 0.40\pm 0.03$ when $k_3\ne 0$ and $1/z = 0.26\pm 0.01$ when $k_3=0$ in model A; $1/z = 0.41 \pm 0.02$ when $k_3\ne 0$ and $1/z = 0.27\pm 0.02$ when $k_3=0$ in model B. In model A, $k_3=3\cdot 10^3$ when nonzero, $k_4=2\cdot 10^3$, $N=10^3$, averaging over 50 different initial conditions. In model B, $F=50$, $k_3=3\cdot 10^3$ when nonzero, $k_4=1.8\cdot 10^6$, $N=10^4$, with one random realization shown. The characteristic timescale $\tau_*$ is $\tau_* = 2000/\varepsilon_*$ in the upper panel, whereas $\tau_* = 1/\varepsilon_*$ in the lower panel, with $\varepsilon_*$ the energy density. \textbf{Bottom:}  Scaling collapse of the equal-time momentum correlation function in model A for $1/z \approx 0.4$.  Shown in the inset is the raw equal-time momentum correlation function without rescaling.  
}
\label{fig:z4}
\end{figure}

%the linearized hydrodynamic theory predicts that \begin{equation}
%    \langle p_k(t)p_{-k}(t)\rangle = \langle p_kp_{-k}\rangle_{\mathrm{eq}} + C_k \mathrm{e}^{-\gamma k^4 t},
%\end{equation} 
%where $C_k$ and $\gamma$ are constants.  We have found it easiest to identify the maximum of $\langle p_k(t)p_{-k}(t)\rangle$ as a function of $k$ (call it $k_*$) at each time $t$.  If the dynamics have critical exponent $z$, we predict \begin{equation}
%    k_*(t) \propto t^{-1/z};
%\end{equation}
Next, we study the thermalization time scale of each model. Choosing random initial conditions, higher-momenta modes will thermalize first, so that $\langle p_{-k}(t)p_k(t)\rangle$ will develop a maximum at some momentum $k_*$. For a dispersion relation with $\text{Im}\,\omega\propto k^z$, the thermalization time of a mode with momentum $k$ scales as $k^{-z}$, which will yield
\be k_*(t)\propto t^{-1/z};\ee
note that $z=4$ in linearized hydrodynamics.  Figure \ref{fig:z4} shows numerical simulations in both models, which demonstrate that $z\approx 4$ when $k_3=0$, but that $z\approx 2.5$ when $k_3\ne 0$.  Crucially, we observe both (\emph{i}) a large scale deviation from $z=4$, which is compatible with our crude estimates for $z$ at the non-equilibrium fixed point, and (\emph{ii}) much weaker instability (in fact, we did not numerically detect one unambiguously) when $k_3=0$.   Moreover, our data exhibits the strongest scaling collapse when $z\approx 2.5$ (when $k_3\ne 0$).  This constitutes strong numerical evidence for the existence of \emph{the same} hydrodynamic fixed point, arising both in Models A and B.   Remarkably, the observed value of $z$ is extremely close to our simplistic estimate of 2.5.

\begin{figure}
\includegraphics[scale = 1.0]{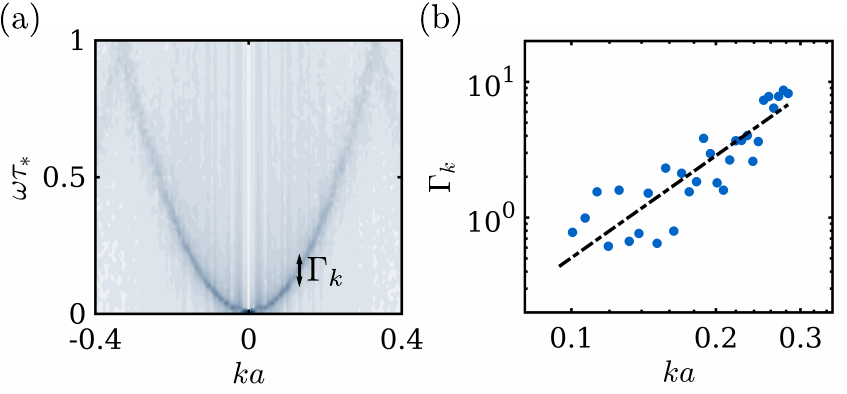}
\caption{(a) Absolute value of the temporal Fourier transform of $\langle P_{-k}(0)P_{k}(t)\rangle$ showing the quadratic dispersion of a propagating mode in model A. (b) Linewidth of the quadratic excitations in panel (a) as a function of momentum. Shown with dashed-dotted lines is the $k^{2.5}$ fit.}
\label{fig:magnon}
\end{figure}

In Model A, we have also studied the fate of the magnon-like sound mode by studying the Fourier transform of unequal time correlation functions $\langle p_{-k}(0)p_k(t)\rangle$.  Figure \ref{fig:magnon} shows that this correlation function is sharply peaked near $\omega = c k^2$, with a magnon decay rate consistent with $z\approx 2.5$.  This suggests that the real part of the dispersion relation remains quadratic at the dipole-conserving KPZ fixed point, and is consistent with our assumption that the densities do not pick up anomalous exponents at this new fixed point.

We remark that the numerical detection of anomalous transport scaling can be sometimes quite subtle. For example, it has long been known that energy transport in one-dimensional ``standard'' (non-fractonic) hydrodynamics is anomalous \cite{PhysRevLett.78.1896}; however, this statement has been subject to relatively recent debates, where certain models were observed to display ordinary diffusion \cite{PhysRevE.82.061118,PhysRevE.85.060102}. While the general consensus is now that energy transport in these systems is always anomalous \cite{PhysRevE.90.012124}, these works instruct us that unambiguous determinations of non-equilibrium critical exponents may be quite non-trivial.  
The relatively weak mixing of energy density with momentum density within these one-dimensional models is consistent with the fact that our model A appears to access the dipole-KPZ fixed point at numerically accessible time scales, even without any noise.
%: see Appendix \ref{app:modelA} for more discussion.

\section{Outlook}We have discovered a new phase of matter which is undriven, yet out-of-equilibrium.  Our construction was inspired by the physics of fractons, which were originally devised \cite{chamon2005quantum} to protect quantum information against thermalization, but have since revealed deep connections between quantum information, condensed matter physics, quantum field theory, and (due to this work) non-equilibrium statistical mechanics.  Although we have focused on the hydrodynamics of a dipole-conserving fluid here, we anticipate infinitely many additional non-equilibrium universality classes arising from the consideration of higher multipole conservation laws \cite{fractonhydro},  subsystem symmetries \cite{fractonhydro,IaconisVijayNandkishore,knap2021}, or explicit/spontaneous symmetry breaking.  We look forward to the systematic classification of fracton-inspired non-equilibrium universality classes, and hope for their ultimate discovery in experiment.

\section*{Acknowledgements}We acknowledge useful discussions with Luca Delacr\'etaz, Bertrand Halperin, Kristan Jensen, Rahul Nandkishore, Leo Radzihovsky and Dam Thanh Son.  

This work was supported by the Department of Energy through Award DE-SC0019380 (PG), the Simons Foundation through Award No. 620869 (PG), the Gordon and Betty Moore Foundation's EPiQS Initiative via Grants GBMF4302 and GBMF8686 (JFRN), and by Research Fellowships from the Alfred P. Sloan Foundation under Grant FG-2020-13615 (PG) and FG-2020-13795 (AL).  
%AL was partially supported by a Research Fellowship from the Alfred P. Sloan Foundation through Grant . JFRN acknowledges the Gordon and Betty Moore Foundation's EPiQS Initiative through Grant GBMF4302 and GBMF8686. 
%JG and AL were partially supported by the Gordon and Betty Moore Foundation's EPiQS Initiative.  (award number will come by v2...)

\emph{Note Added.}---While preparing this work, the preprint \cite{grosvenor2021hydrodynamics}, which discusses non-dissipative response in a fluid with conserved dipole and momentum, was posted.

%COMMENT OUT FOR ARXIV

%\emph{Author Contributions.}---PG and AL developed the theoretical framework and classical models.  JRN simulated Model A.  JG simulated Model B.  PG and AL wrote the manuscript with input from all authors.

%\emph{Conflicts of Interest.}---The authors declare no conflict of interest.

\appendix

\section{Zeroth order hydrodynamics}\label{app:zero}
In this appendix, we derive the zeroth order hydrodynamics.  Our starting point is (\ref{2ndla0}).  First collecting the terms with time derivatives, we find \begin{align}
    T\partial_t s &= -V^i \partial_t \pi^i - \mu \partial_t n \notag \\
   T\left(\frac{\partial s}{\partial n} \partial_t n + \frac{\partial s}{\partial (\partial_k v^l)} \partial_t \partial_k v_l\right) &= -V^i \partial_t (nv^i) - \mu \partial_t n \notag \\
 \left(T\frac{\partial s}{\partial n}+ V^i v^i + \mu\right)\partial_t n  + \psi_{kl}\partial_t \partial_k &v_l + nV^i \partial_t v^i  = 0.
\end{align}
Since the first term must vanish, we conclude that the chemical potential is given by \begin{equation}
    \mu = -T\frac{\partial s}{\partial n} - V^iv^i.\label{chempot}
\end{equation}
For the second term, we integrate by parts to find \begin{equation}
   \partial_k\left( \psi_{kl}\partial_t  v_l\right) + \left(nV^i - \partial_k \psi_{kl}\right) \partial_t v^i = 0.
\end{equation}
Setting the second term to zero implies the first equation in (\ref{const2}).  

The remaining divergence term can be absorbed into the entropy current, and indeed we now turn to the evaluation of the entropy current by studying the spatial derivative terms in (\ref{2ndla0}): \begin{align} \label{eq:AJS}
    &T\partial_i J_S^i + \partial_k\left( \psi_{kl}\partial_t  v_l\right) = -V^i \partial_j T^{ji} - \mu \partial_i J^i \notag \\
    &= -V^i \partial_j T^{ji} + \left(T \frac{\partial s}{\partial n} + V^k v^k\right)\partial_i \left(nV^i + \tilde J^i\right) .
\end{align}
In the second line above, we have plugged in the ansatz $J^i = nV^i + \tilde J^i$.  Ultimately we will show that $\tilde J^i = 0$ at ideal fluid level, so in the equations that follow we will drop this term to avoid clutter.  With a bit of hindsight, we claim that \begin{equation}
    J_S^i = n\frac{\partial s}{\partial n} V^i - \psi_{ij}\partial_t \psi_j. \label{eq:JS}
\end{equation}
The latter term clearly cancels out a term in the first line of (\ref{eq:AJS}); as for the former, observe that if we plug (\ref{eq:JS}) into (\ref{eq:AJS}), we find \begin{align}
    -V^i &\partial_j T^{ji}+ V^k v^k \partial_i (nV^i) =  TnV^i \partial_i \frac{\partial s}{\partial n} \notag \\
    &= TV^i\partial_i \left(n\frac{\partial s}{\partial n}\right) - TV^i \frac{\partial s}{\partial n}\partial_i n \notag \\
    &= TV^i\partial_i \left(n\frac{\partial s}{\partial n} - s \right)+ TV^i  \frac{\partial s}{\partial \partial_k v^l} \partial_i \partial_k v^l \notag \\
    &= TV^i\partial_i \left(n\frac{\partial s}{\partial n} - s \right) + V^i \partial_k \left(\psi_{kl}\partial_i v^l\right) \notag \\
    &\; \;\;\;\;\;\;\; - nV^i  V^l\partial_i v^l.
\end{align}
This equation is satisfied if \begin{equation}
    T^{ji} = T\left(s-n\frac{\partial s}{\partial n}\right)\delta^{ji} + nV^jv^i-\psi_{jk}\partial_i v_k.
\end{equation}

%\begin{subequations}\begin{align}
%     J_S^i &= s V^i+\Delta J_S^i,\\ 
%     T^{ij} &= p \delta^{ij} + V^i \pi^j+\Delta T^{ij}, \\
%     J^i &= n V^i +\Delta J^i,
%    \end{align}\end{subequations}  

%where $\Delta J_S^i,\Delta T^{ij},\Delta J^i$ vanish for a non-dipole conserving fluid with thermodynamics $s=s(\pi^i,n)$, and are local expressions in $v_i$ and $n$ to be determined. Plugging in (\ref{2ndla0}), we find \PG{(should add the explicit derivation of the below in the appendix?)} Yes, do this here now!!
%\begin{subequations}\begin{gather} T^{ij}=p \delta^{ij}+ V^i \pi^j-\psi_{ik}\p_j v_k,\qquad J^i= n V^i\label{const1} \\
%V^i=\frac 1n \p_j \psi_{ji},\qquad \mu=-T\frac{\p s}{\p n}-V^i v^i\\
%J_S^i=n\frac{\p s}{\p n}V^i-\psi_{ij}\p_t v_j,\qquad p=Ts+\mu n+V^i \pi^i \ ,\label{const3}
%\end{gather}\end{subequations}

%and where $\frac{\p s}{\p n}$ is performed at fixed $v^i=\frac{\pi^i}n$. As we show in Appendix \ref{app:sym}, while it is not manifest from (\ref{const1}), the stress tensor $T^{ij}$ is symmetric \emph{up to total derivatives}. 
%The first equation in (\ref{const2}) precisely recovers (\ref{dipfl}). 

Let us confirm that this ideal stress tensor $T^{ij}$ is symmetric in the indices $i,j$ up to total derivatives. To this aim, consider a constant antisymmetric matrix $\Omega_{ij}$ and write
\be\begin{split} &\int \mathrm{d}^d x T^{ij}\Omega_{ij}\\
&=-T\int \mathrm{d}^dx \left(\frac{\p s}{\p(\p_k v_i)} \p_k v_j+\frac{\p s}{\p(\p_i v_k)}\p_j v_k\right)\Omega_{ij}\\
&=-T \int \mathrm{d}^dx \frac{\p s}{(\p_i v_j)}\delta(\p_i v_j)\end{split}\ee
where $\delta(\p_i v_j)=\left(\Omega_{jk}\p_i v_k+\Omega_{ik}\p_k v_j\right)$ is the variation of $\p_i v_j$ with respect to the infinitesimal rotation $x^i\to x^i+\Omega^{ij}x^j$. Since the entropy density $s$ is invariant under rotations, the above expression vanishes.  Hence, $T_{ij}$ is symmetric, up to total derivative contributions.

\section{Dissipative hydrodynamics}\label{app:const}
Introduce higher-derivative corrections to the currents: $T^{ij}=T^{ij}_{(0)}+T^{ij}_{(1)}$, $J^{ij}=J^{ij}_{(0)}+J^{ij}_{(1)}$, and $J^i=J^i_{S(0)}+J^i_{S(1)}$ where the quantities with subscript $(1)$ denote higher-derivative corrections. Using the conservation laws (\ref{eom1}) together with thermodynamic relations (\ref{const2}) and (\ref{chempot}), expand 
\begin{align} T\p_t s=&T\frac{\p s}{\p n}\p_tn+\psi_{ij}\p_t\p_i v_j\nonumber\\
=&\mu\p_i\p_j J^{ij}+V^j\p_i T^{ij}+\p_i(\psi_{ij}\p_tv_j)\nonumber\\
=&\p_i(\mu \p_j J^{ji}-J^{ij}\p_j\mu+V^j T^{ij}\psi_{ij}\p_t v_j)\nonumber\\
&+J^{ij}\p_i\p_j\mu-T^{ij}\p_i V_j.\end{align}
Eq. (\ref{entp}) then gives
\begin{gather} 0\leq
\p_i(J_{S(1)}^i+\mu \p_j J_{(1)}^{ji}+V^j T_{(1)}^{ij}\psi_{ij}(\p_t v_j)_{(1)}-J_{(1)}^{ij}\p_j\mu)
\nonumber\\+J_{(1)}^{ij}\p_i\p_j\mu-T_{(1)}^{ij}\p_i V_j,\label{entp1}\end{gather}
where the leading derivative terms cancel out due to (\ref{2ndla0}), and where $(\p_t v_j)_{(1)}$ denotes the dissipative contribution to $\p_t v_j$. The most general expressions for the dissipative stress tensor and charge current are
\begin{subequations}\label{con2}\begin{align}\label{con2a} -T^{ij}_{(1)}&=\eta_1^{ijkqpl}\p_k\p_l\p_p v_q+\alpha^{ijkl}\p_k\p_l \mu\\
J^{ij}_{(1)}&=\kappa_1^{ijkqpl}\p_k\p_l\p_p v_q+C^{ijkl}\p_k\p_l \mu\label{con3}\ ,\end{align}\end{subequations}
where $\eta_1^{ijkpql},\dots$ are regarded as constant tensors, and where we only kept terms that are linear in $v_i$ and $\mu$ as these are sufficient to capture the anomalous scalings: any nonlinear dissipative term will be irrelevant. The tensor $\eta_1^{ijkqpl}$ has the general decomposition
\be\begin{split}\label{eta1aa} \eta_1^{ij(kqp)l}=&b_1\delta^{ij}\delta^{(kp}\delta^{q)l}+b_2\delta^{i(k}\delta^{|j|p}\delta^{q)l}\\
&+b_3\delta^{jl}\delta^{i(k}\delta^{pq)}+b_4\delta^{il}\delta^{j(k}\delta^{pq)},\end{split}\ee
where $A^{(k|r_1r_2\dots|pq)}$ denotes total symmetrization with respect to the indices $k,p,q$. From (\ref{eta1}) and (\ref{aijkl}) below, we also have
\be\begin{split}\label{etaa} \eta^{ij(k|r|}a^{p|r|q)l}=&\left(\zeta-\frac 2d\eta\right)\left(a_1+2\frac{d-1}d a_2\right)\delta^{ij}\delta^{(kp}\delta^{q)l}\\
&+2\eta\left(a_1+\frac{d-2}d a_2-a_3\right)\delta^{i(k}\delta^{|j|p}\delta^{q)l}\\
&+(\eta+\bar\eta)(a_2+a_3)\delta^{i(k}\delta^{pq)}\delta^{lj}\\
&+(\eta-\bar\eta)(a_2+a_3)\delta^{il}\delta^{j(k}\delta^{pq)}.\end{split}\ee
Eqs. (\ref{eta1aa}) and (\ref{etaa}) have the same tensor structure. From this and (\ref{Vmp}) we conclude that, by a suitable choice of coefficients, the first term in (\ref{con2a}) can be written as
\be \eta_1^{ijklpq}\p_k\p_l\p_p v_q=\eta^{ijkl}\p_k V_l\ee
A similar conclusion applies to the first term in (\ref{con3}). This allows us to write the currents in terms of $V^i$:
\begin{subequations}\label{con1ap}\begin{align} -T^{ij}_{(1)}&=\eta^{ijkl}\p_k V_l+\alpha^{ijkl}\p_k\p_l \mu\\
J^{ij}_{(1)}&=\kappa^{ijkl}\p_k V_l+C^{ijkl}\p_k\p_l \mu,\end{align}\end{subequations}
where
\begin{subequations}
\begin{align}
\eta^{ijkl}&=\zeta \delta^{ij}\delta^{kl}+2\eta\delta^{i<k}\delta^{l>j}+2 \bar\eta \delta^{i[k}\delta^{l]j}\label{eta1}\\
\kappa^{ijkl}&=\kappa_1 \delta^{ij}\delta^{kl}+2\kappa_2\delta^{i<k}\delta^{l>j}+2 \kappa_3 \delta^{i[k}\delta^{l]j}\\
\alpha^{ijkl}&=\alpha_1\delta^{ij}\delta^{kl}+2\alpha_2\delta^{i<k}\delta^{j>l}\\
C^{ijkl}&=C_1\delta^{ij}\delta^{kl}+2C_2\delta^{i<k}\delta^{j>l}
,\end{align}\end{subequations}
where $A^{<ij>}=\frac 12(A^{ij}+A^{ji}-\frac 1d \delta^{ij}A^{kk})$ and $A^{[ij]}=\frac 12(A^{ij}-A^{ji})$ denote traceless symmetrization and anti-symmetrization with respect to indices $i,j$, respectively. The coefficients in (\ref{con1ap}) can be regarded as a matrix acting on the vector $(\p_i V_j,\p_i\p_j \mu)$. Choosing $J_{S(1)}^i$ so that the total derivative on the right-hand side of (\ref{entp1}) vanishes then implies positive semi-definiteness of such transport matrix. Moreover, Onsager's principle implies that this matrix should be symmetric. The last term in (\ref{eta1}) would contribute to entropy production through rigidly rotating the fluid as $\mathrm{d}S\sim \bar\eta (\p_{[i}V_{j]})^2$; this cannot happen for rotationally invariant fluids, and we shall set it to zero. Similarly, by coupling the system to a suitable background higher-rank gauge field, one can infer that $\kappa_3$ also vanishes. We shall explicitly derive these constraints in \cite{future}. Putting all the constraints together, we then find
\be\begin{gathered}\label{constr}
\zeta,\eta,C_1,C_2\geq 0,\qquad \alpha_1=\kappa_1,\quad \alpha_2=\kappa_2\\
\alpha_1^2\leq \zeta C_1,\qquad \alpha_2^2\leq \eta C_2.
\end{gathered}\ee

\section{Linearized hydrodynamics} \label{app:linearized}
We now analyze the dispersion relations of the linearized hydrodynamics around a homogeneous background charge density $n_0$, ignoring the effects of fluctuations. Taking $n=n_0+\delta n$, where $\delta n$ and $v_i$ are regarded as small, expand
\be\label{entdens} s=-\frac 12 a^{ijkl}\p_i v_j \p_k v_l-\frac{\mu_0}T\delta n-\frac 1{2\chi T}\delta n^2,\ee
where the tensor $a^{ijkl}=a^{klij}$ has the following general decomposition:
\be\label{aijkl} a^{ijkl}=a_1 \delta^{ij}\delta^{kl}+2a_2\delta^{i<k}\delta^{l>j}+2 a_3 \delta^{i[k}\delta^{l]j}\ee
In (\ref{entdens}), $\mu_0=\mu(n_0)$ is the chemical potential evaluated on the background charge density, and $\chi=\frac{\p n}{\p \mu}$ is charge susceptibility. Thermodynamic stability requires that $a_1,a_2,a_3,\chi\geq 0$. The thermodynamic relations in Eqs. (\ref{const2}),(\ref{chempot}) then give
\be\label{Vmp}\begin{gathered} V^i=-\frac T{n_0^2}a^{jikl}\p_j\p_k\pi_l,\quad \delta\mu=\chi^{-1}\delta n\\
p=\mu_0 n_0+\chi^{-1}n_0\delta n\end{gathered}\ee
where $\delta\mu=\mu-\mu_0$, and the ideal part of the currents read
\begin{subequations}
\begin{align}
    T_{(0)}^{ij}&=(\mu_0n_0+n_0\chi^{-1}\delta n)\delta^{ij}, \\
    J^{ij}_{(0)}&=-\frac T{n_0}a^{ijkl}\p_k\pi_l,
\end{align}
\end{subequations}
while the dissipative contributions $T^{ij}_{(1)}$ and $J^{ij}_{(1)}$ are precisely given by (\ref{con1}). Plugging these in the conservation equations (\ref{eom1}) yields
\begin{subequations}
\begin{align}
    &\p_t\delta n-\frac{aT}{n_0} \p^2\p_i\pi_i+\frac C\chi \p^4\delta n-\frac {aT}{n_0^2}K\p^4\p_i\pi_i=0\\
    &\p_t\pi^i+\frac{n_0}\chi \p_i\delta n+\frac {aT}{n_0^2}\Gamma_1\p^4\pi_i
    +\frac {aT}{n_0^2}\Gamma_2 \p^2\p_i \p_j \pi_j\nonumber\\
    &\hspace{4.19cm}-\frac A\chi \p_i \p^2\delta n=0
\end{align}
\end{subequations}
where 
\begin{subequations}
\begin{align}
    a&=a_1+2\frac{d-1}da_2, \\
    C&=C_1+2\frac{d-1}d C_2, \\
    K&=\kappa_1+2\frac{d-1}d\kappa_2, \\
    \Gamma_1&=\frac{a_2+a_3}{a}\eta, \\
    \Gamma_2&=\zeta+\left(2\frac{d-1}da-\frac 1{d^2}a_2-a_3\right)\frac{\eta}{a}, \\
    A&=\alpha_1+2\frac{d-1}d\alpha_2.
\end{align}
\end{subequations}
These equations lead to the following dispersion relations:\begin{subequations}
\begin{align} \omega& =\pm \sqrt{\frac{aT n_0}\chi}k^2-\frac{\mathrm{i}}{2}\left(\frac C\chi+\frac {aT}{n_0^2}(\Gamma_1+\Gamma_2)\right)k^4\\
\omega&=-\mathrm{i}\frac {aT}{n_0^2}\Gamma_1 k^4
\end{align}\end{subequations}
where the two modes in the first line arise from the coupling between $\delta n$ and the longitudinal component of momentum $\p_i \pi_i$, while the modes in the second line have multiplicity $d-1$ and are associated to the transverse components of momentum $\p^{-2}(\pi_i-\p_i\p_j\pi_j)$. 
%[Paolo: Note $\Gamma_1\propto a_2-a_3$, which seems to imply $a_3\leq a_2$. I have no idea how to derive this bound (even using the EFT)).]

\section{Equations of motion with relevant nonlinearities}\label{app:expl}
Here we give explicit expressions for various quantities, valid beyond linear response. The input is the entropy density
\be s=-\frac 12 a^{ijkl}\p_i v_j \p_k v_l +\bar s(n),\ee
whose form is sufficient to capture all the relevant nonlinearities, and $\bar s(n)$ is a generic function of $n$. Using (\ref{const1})-(\ref{const4}) we have
\be\begin{gathered} \psi_{ij}=-Ta^{ijkl}\p_k v_l,\quad nV^i=-T a^{jikl}\p_j\p_k v_l\\
p=-\frac T2 a^{ijkl}\p_i v_j\p_k v_l+\bar p(n)\end{gathered}\ee
where $\bar p =T(\bar s-n\p_n \bar s)$. The currents are
\begin{align}
    T^{ij}=&\left(\bar p-\frac T2 a^{klpq}\p_k v_l\p_p v_q\right)\delta^{ij}-T a^{pjkl}\p_p \p_kv_lv^i\nonumber\\
    &+Ta^{jkpl}\p_p v_l \p_i v_k-\eta^{ijkl}\p_k V_l-\alpha^{ijkl}\p_k\p_l \mu+\tau^{ij}\nonumber\\
    J^{ij}=&-Ta^{ijkl}\p_k v_l+C^{ijkl}\p_k\p_l\mu+\alpha^{ijkl}\p_k V_l+\xi^{ij},
\end{align}
where we already accounted for the constraints discussed around (\ref{constr}). It will be convenient to express the conservation equations in terms of $v_i=\frac{\pi_i}n$, for which we have
\be \p_t v^i+\frac 1n\left(\p_j T^{ji}-v^i \p_k \p_j J^{jk}\right)=0.\ee
Plugging in the above expressions and substituting (\ref{aijkl}) for $a^{ijkl}$, we find
\begin{subequations}\label{eomnl}\begin{align} &\p_t v_i+\frac 1n \p_i \bar p+\frac {2T}n(a-a_2-a_3)\p_j\p_k v_k\p_{[i}v_{j]}\nonumber\\
&+\frac{2T}n(a_2+a_3)\p_j^2 v_k\p_{[i}v_{k]}+\frac {aT}{n^2}\Gamma_1\p^4v_i\nonumber\\
&+\frac {aT}{n^2}\Gamma_2 \p^2\p_i \p_j v_j
-\frac A{n\chi} \p_i \p^2 n+\frac 1n\p_j\tau^{ji}=0\label{mom2}\\
&\p_t n-aT \p^2\p_iv_i+\frac C\chi \p^4 n-\frac {aT}{n}A\p^4\p_iv_i+\p_i\p_j \xi^{ij}=0
\end{align}\end{subequations}
where we neglected nonlinear dissipative terms, since they are irrelevant.

Expanding
\be \bar s(n)=-\frac{\mu_0}T\delta n-\frac 1{2\chi T}\delta n^2-\frac \lambda{6 T}\delta n^3-\frac {\lambda_1}{12 T}\delta n^4,
\ee
we find that the contribution from $\bar p$ in (\ref{mom2}) is
\be \frac 1n\p_i \bar p=\frac 1\chi \p_i\delta n+\lambda\delta n\p_i\delta n+\lambda'\delta n^2\p_i\delta n,\ee
where $\lambda'=\lambda_1-\frac 1{\chi n_0^2}$.

The evaluation of the dynamical exponent $z$ is not amenable to perturbation theory. One can verify this by evaluating the Kubo formula for frequency-dependent transport; for example for the viscosity 
\be\begin{split} \eta(\omega)&\sim \int \mathrm{d}t \mathrm{e}^{-\mathrm{i}\omega t}\int \mathrm{d}^d x\langle T^{xx}(t,\vec x)T^{xx}(0,0)\rangle\\
&\sim\int \mathrm{d}t \mathrm{e}^{-\mathrm{i}\omega t}\int \mathrm{d}^d x\langle (\delta n(t,\vec x))^2(\delta n(0,0))^2\rangle\\
&\sim \int \mathrm{d}^d k\mathrm{d}t \mathrm{e}^{-\mathrm{i}\omega t}\mathrm{e}^{-2\gamma k^4t}\sim \omega^{\frac{d-4}4},\end{split}\ee
where we only considered the nonlinearity $T^{ij}\sim \delta n^2\delta^{ij}$ and assumed factorization of the 4-point function of $\delta n$. For simplicity we assumed absence of propagating modes, so that $\langle n(t,k)n(0,-k)\rangle\sim \mathrm{e}^{-\gamma k^4 t}$. We then see that the correction to the mean-field viscosity is divergent as $\omega\to 0$, making perturbation theory ill-defined.

\section{Stochastic Model A} \label{app:modelA}
\begin{figure}
\includegraphics[scale = 1.0]{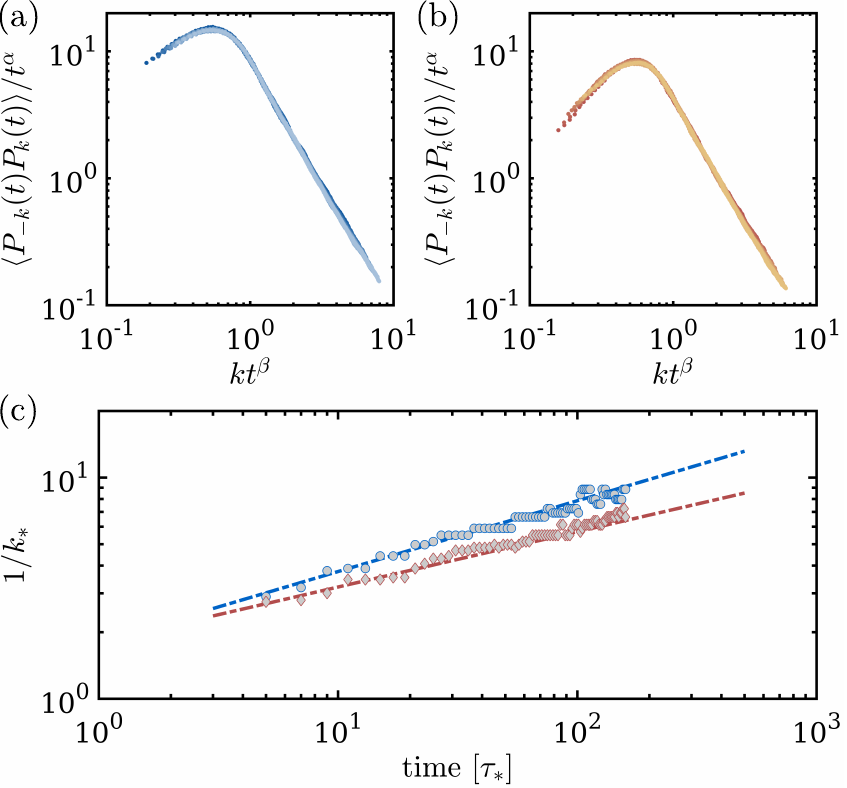}
\caption{Scaling of momentum correlation for the stochastic model A. Shown in panels (a-b) are the self-similar collapse of the correlations for (a) $\nu = 0.1$, and (b) $\nu = 1$. Shown in panel (c) is the position of the peak of the distribution as a function of time for (a) and (b). The linear fitting yields $1/z = 0.31 \pm 0.02$ for $\nu = 0.1$, and $1/z = 0.26\pm 0.02$ for $\nu =1$. }
\label{fig:noise}
\end{figure}

We now overview a stochastic version of model A, given by the equations:
\begin{subequations}
\begin{align} &\p_t p_n-V'(x_{n-1}-x_n)+V'(x_n-x_{n+1}) \notag \\
&+\nu (2\p_t x_n -\p_tx_{n-1}-\p_tx_{n+1})+\xi_n-\xi_{n-1}=0\\
&\p_t x_n-(2 p_n-p_{n+1}-p_{n-1}) \notag \\
&-\nu(2\p_t p_n-\p_tp_{n-1}-\p_tp_{n+1})+\zeta_i-\zeta_{i-1}=0
\end{align}\end{subequations}
where $\nu$ is a friction coefficient, and the noise correlations are chosen so that the fluctuation-dissipation theorem is satisfied:
\be \langle \xi_n(t)\xi_m(0)\rangle=\langle\zeta_n(t)\zeta_m(0)\rangle=\nu T \delta_{mn}\delta(t).\ee
Setting $\nu=0$, one recovers the Hamiltonian model A. The model above can serve to verify that the scaling exponent $z$ remains smaller than 4 even after energy conservation is broken, which happens, at sufficiently long times, as soon as $\nu$ is turned on. Figure \ref{fig:noise} shows that, for $\nu$ not too large the scaling exponent is still anomalous, thus corroborating the fact that $z<4$ independently of the existence of a diffusive energy conservation law.   We emphasize that in the true thermodynamic limit, theories with any amount of noise are expected to flow to the same fractonic KPZ theory -- the fact that our estimate of $z$ changes with $\nu$ is a consequence of finite size effects.

%\begin{figure}
%\includegraphics[scale = 1.0]{./figures/energy.pdf}
%\caption{Fourier transform of the energy-energy correlation function $\langle E_{-k}(0)E_{k}(t))\rangle$. Energy transport is primarily diffusive, but there is a very weak coupling with the quadratic propagating mode (note the faint parabola in the spectrum). The right pannel shows a linecut of the FT. }
%\label{fig:texturenumber}
%\end{figure}

%\begin{figure}
%\includegraphics[scale = 1.0]{./figures/thermalization.pdf}
%\caption{For strong non-linearities, energy looks thermal in O(1) timescales, much faster than the characteristic timescales in which the peak in figure 1 moves. Shown is the distribution of local energies $E_{i}$. }
%\label{fig:texturenumber}
%\end{figure}

\section{Linearized dynamics in model B} \label{app:modelB}
Here we show that the linearized equations in model B exhibit fast Bloch oscillations superimposed on top of magnon-like
hydrodynamic modes. 
Consider the Hamiltonian of model B with ${V(x) = \frac{1}{2} x^2}$:
\be H = \sum_{i=1}^N \left(-\cos p_i - F x_i\right) + \frac{1}{2} (x_i-x_{i-1})^2 .\ee
We observe that the nonlinear equations of motion:
\begin{equation}\begin{aligned} 
&\partial_t x_i = \sin p_i  \\
&\partial_t p_i = F + 2x_i - x_{i+1} - x_{i-1},
\end{aligned}\end{equation}
admit the following exact solution corresponding to collective Bloch oscillations: \begin{subequations}
\begin{align}
    x^0_i(t) &= \frac{1-\cos (Ft)}{F}, \\
    p^0_i(t) &= Ft.
\end{align}
\end{subequations}
Now consider perturbations around this solution:  $\delta x_i = x_i - x_i^0$ and $\delta p_i = p_i - p_i^0$.  The linearized equations of motion for the perturbations are
\begin{equation}\label{eomb}\begin{aligned} 
&\partial_t \delta x_k = \cos Ft \ \delta p_k , \\
&\partial_t \delta p_k = (2-2\cos k) x_k,
\end{aligned}\end{equation}
where ${\delta x_k}$ and ${\delta p_k}$ are the discrete Fourier transform of ${\delta x_i}$ and ${\delta p_i}$.
When $k\ll 1$ (the long wavelength limit), (\ref{eomb}) implies
\begin{equation}
\partial^2_t \delta p_k = k^2 \cos{Ft}\ \delta p_k,
\end{equation}
which is the well-studied Mathieu equation. 

The general form of the Mathieu equation is an ordinary differential equation with real coefficients:
\begin{equation}\label{mathieu}
u^{\prime \prime}+(a-2 q \cos 2 \tau) u=0.
\end{equation}
In our case, ${a = 0}$, ${\tau = Ft/2}$, ${q = 2k^2/F^2 \ll 1}$. By Floquet’s theorem, (\ref{mathieu}) has solutions of the form ${\mathrm{e}^{\mathrm{i}\nu \tau} \Phi(\tau)}$ and ${\mathrm{e}^{-\mathrm{i}\nu \tau} \Phi(-\tau)}$, where $\Phi(\tau)$ is periodic with period $\pi$.  Assuming ${\nu}$ is not an integer, ${\mathrm{e}^{\mathrm{i}\nu \tau} \Phi(\tau)}$ and ${\mathrm{e}^{-\mathrm{i}\nu \tau} \Phi(-\tau)}$ are linearly independent, so we can expand ${u(\tau)}$ as
\begin{equation}\label{mathieuexp}
u(\tau)=A_{1} \mathrm{e}^{\mathrm{i}\nu \tau} \sum_{n=-\infty}^{\infty} c_{2 n} \mathrm{e}^{2 n \mathrm{i} \tau}+A_{2} \mathrm{e}^{-\mathrm{i}\nu \tau} \sum_{n=-\infty}^{\infty} c_{2 n} \mathrm{e}^{-2 n \mathrm{i} \tau}.
\end{equation}
Substituting (\ref{mathieuexp}) in (\ref{mathieu}) gives a recurrence relationship for ${c_{2n}}$:
\begin{equation}\label{recr}
\gamma_{n}(\nu) c_{2 n-2}+c_{2 n}+\gamma_{n}(\nu) c_{2 n+2}=0
\end{equation}
where
\begin{equation}
\gamma_{n}(\nu)=\frac{q}{(2 n-\nu)^{2}-a}=\frac{2k^2}{F^2(2n-\nu)^2}.
\end{equation}
The equations (\ref{recr}) can be written as a matrix equation ${M_{mn}c_{n} = 0}$, with repeated indices summed over, and with tridiagonal matrix ${M_{mn} = \delta_{mn} + \delta_{m,n-1} \gamma_{m}+ \delta_{m,n+1} \gamma_{m}}$. The exponent ${\nu}$ is fixed by the condition $\det(M)=0$. 

Assuming ${\nu \ll 1}$, we have the relations ${\gamma_{0}(\nu) \gg \gamma_{n}(\nu)}$ and ${\gamma_{n}(\nu) \ll 1}$ for any ${n \neq 0}$. So ${M}$ is now a tridiagonal matrix whose diagonal elements are 1 and all off-diagonal elements other than ${M_{0,1}}$ and ${M_{0,-1}}$ are perturbatively small. As a result, we can estimate the value of ${\nu}$ by only considering the matrix elements ${M_{mn}}$ for ${m}$, ${n = -1,\ 0,\ 1}$:
\begin{equation}
\det(M) \approx \det \left(\begin{array}{ccc} 1 & \frac{q}{4} & 0 \\ \frac{q}{\nu^2} & 1 & \frac{q}{\nu^2} \\ 0 & \frac{q}{4} & 1 \end{array}\right) = 1-\frac{q^2}{2\nu^2} = 0
\end{equation}
We get ${\nu = \frac{q}{\sqrt{2}} = \frac{\sqrt{2}k^2}{F^2} \ll 1}$. Since now ${\frac{q}{\nu^2} = \frac{2}{q} \gg 1}$, we have ${\left|c_0\right| \gg \left|c_{\pm 2}\right| \gg \left|c_{\pm2n}\right|}$ for any ${n > 1}$, so approximately ${c_2 = c_{-2} \approx -\frac{q}{4}c_0}$. 

Assuming the initial condition ${\delta p_k(0) = 0}$, we have ${A_1 = -A_2}$, so
\begin{align}
\delta p_k(t) &\approx \left(A_{1} \mathrm{e}^{\mathrm{i}\nu \tau} + A_{2} \mathrm{e}^{-\mathrm{i}\nu \tau}\right) \left(c_0 +c_{\pm 2} \mathrm{e}^{2 \mathrm{i} \tau} + c_{\pm 2} \mathrm{e}^{-2 \mathrm{i} \tau} \right) \notag \\
&= A \sin{\nu \tau}\left(1-\frac{q}{2}\cos{2\tau} \right) \notag \\
&= A \sin{\frac{k^2 t}{\sqrt{2}F}} \left(1-\frac{k^2}{F^2}\cos{Ft} \right)
\end{align}
We can also find ${\delta x_k}$:
\begin{equation}
\delta x_k(t) \approx \frac{A}{\sqrt{2}F} \cos{\frac{k^2 t}{\sqrt{2}F}} + \frac{A}{F} \sin{Ft}\ \sin{\frac{k^2 t}{\sqrt{2}F}}.
\end{equation}
We observe that a fraction of both $\delta p_k$ and $\delta x_k$ oscillate at the frequency $k^2$, confirming our claim that this model has hydrodynamic magnon modes propagating on top of fast Bloch oscillations.

\bibliography{dipolemom}

\begin{thebibliography}{59}%
\makeatletter
\providecommand \@ifxundefined [1]{%
 \@ifx{#1\undefined}
}%
\providecommand \@ifnum [1]{%
 \ifnum #1\expandafter \@firstoftwo
 \else \expandafter \@secondoftwo
 \fi
}%
\providecommand \@ifx [1]{%
 \ifx #1\expandafter \@firstoftwo
 \else \expandafter \@secondoftwo
 \fi
}%
\providecommand \natexlab [1]{#1}%
\providecommand \enquote  [1]{``#1''}%
\providecommand \bibnamefont  [1]{#1}%
\providecommand \bibfnamefont [1]{#1}%
\providecommand \citenamefont [1]{#1}%
\providecommand \href@noop [0]{\@secondoftwo}%
\providecommand \href [0]{\begingroup \@sanitize@url \@href}%
\providecommand \@href[1]{\@@startlink{#1}\@@href}%
\providecommand \@@href[1]{\endgroup#1\@@endlink}%
\providecommand \@sanitize@url [0]{\catcode `\\12\catcode `\$12\catcode
  `\&12\catcode `\#12\catcode `\^12\catcode `\_12\catcode `\%12\relax}%
\providecommand \@@startlink[1]{}%
\providecommand \@@endlink[0]{}%
\providecommand \url  [0]{\begingroup\@sanitize@url \@url }%
\providecommand \@url [1]{\endgroup\@href {#1}{\urlprefix }}%
\providecommand \urlprefix  [0]{URL }%
\providecommand \Eprint [0]{\href }%
\providecommand \doibase [0]{https://doi.org/}%
\providecommand \selectlanguage [0]{\@gobble}%
\providecommand \bibinfo  [0]{\@secondoftwo}%
\providecommand \bibfield  [0]{\@secondoftwo}%
\providecommand \translation [1]{[#1]}%
\providecommand \BibitemOpen [0]{}%
\providecommand \bibitemStop [0]{}%
\providecommand \bibitemNoStop [0]{.\EOS\space}%
\providecommand \EOS [0]{\spacefactor3000\relax}%
\providecommand \BibitemShut  [1]{\csname bibitem#1\endcsname}%
\let\auto@bib@innerbib\@empty
%</preamble>
\bibitem [{\citenamefont {Landau}\ and\ \citenamefont
  {Lifshitz}(1987)}]{landau}%
  \BibitemOpen
  \bibfield  {author} {\bibinfo {author} {\bibfnamefont {L.}~\bibnamefont
  {Landau}}\ and\ \bibinfo {author} {\bibfnamefont {E.}~\bibnamefont
  {Lifshitz}},\ }\href@noop {} {\emph {\bibinfo {title} {Fluid Mechanics}}},\
  \bibinfo {edition} {2nd}\ ed.\ (\bibinfo  {publisher} {Butterworth
  Heinemann},\ \bibinfo {year} {1987})\BibitemShut {NoStop}%
\bibitem [{\citenamefont {Crossley}\ \emph {et~al.}(2017)\citenamefont
  {Crossley}, \citenamefont {Glorioso},\ and\ \citenamefont
  {Liu}}]{Crossley:2015evo}%
  \BibitemOpen
  \bibfield  {author} {\bibinfo {author} {\bibfnamefont {M.}~\bibnamefont
  {Crossley}}, \bibinfo {author} {\bibfnamefont {P.}~\bibnamefont {Glorioso}},\
  and\ \bibinfo {author} {\bibfnamefont {H.}~\bibnamefont {Liu}},\ }\bibfield
  {title} {\bibinfo {title} {{Effective field theory of dissipative fluids}},\
  }\href {https://doi.org/10.1007/JHEP09(2017)095} {\bibfield  {journal}
  {\bibinfo  {journal} {JHEP}\ }\textbf {\bibinfo {volume} {09}},\ \bibinfo
  {pages} {095}},\ \Eprint {https://arxiv.org/abs/1511.03646} {arXiv:1511.03646
  [hep-th]} \BibitemShut {NoStop}%
%%CITATION = ARXIV:1511.03646;%%
\bibitem [{\citenamefont {Haehl}\ \emph {et~al.}(2016)\citenamefont {Haehl},
  \citenamefont {Loganayagam},\ and\ \citenamefont
  {Rangamani}}]{Haehl:2015foa}%
  \BibitemOpen
  \bibfield  {author} {\bibinfo {author} {\bibfnamefont {F.~M.}\ \bibnamefont
  {Haehl}}, \bibinfo {author} {\bibfnamefont {R.}~\bibnamefont {Loganayagam}},\
  and\ \bibinfo {author} {\bibfnamefont {M.}~\bibnamefont {Rangamani}},\
  }\bibfield  {title} {\bibinfo {title} {{The Fluid Manifesto: Emergent
  symmetries, hydrodynamics, and black holes}},\ }\href
  {https://doi.org/10.1007/JHEP01(2016)184} {\bibfield  {journal} {\bibinfo
  {journal} {JHEP}\ }\textbf {\bibinfo {volume} {01}},\ \bibinfo {pages}
  {184}}\BibitemShut {NoStop}%
%%CITATION = ARXIV:1510.02494;%%
\bibitem [{\citenamefont {Jensen}\ \emph {et~al.}(2018)\citenamefont {Jensen},
  \citenamefont {Pinzani-Fokeeva},\ and\ \citenamefont
  {Yarom}}]{Jensen:2017kzi}%
  \BibitemOpen
  \bibfield  {author} {\bibinfo {author} {\bibfnamefont {K.}~\bibnamefont
  {Jensen}}, \bibinfo {author} {\bibfnamefont {N.}~\bibnamefont
  {Pinzani-Fokeeva}},\ and\ \bibinfo {author} {\bibfnamefont {A.}~\bibnamefont
  {Yarom}},\ }\bibfield  {title} {\bibinfo {title} {{Dissipative hydrodynamics
  in superspace}},\ }\href {https://doi.org/10.1007/JHEP09(2018)127} {\bibfield
   {journal} {\bibinfo  {journal} {JHEP}\ }\textbf {\bibinfo {volume} {09}},\
  \bibinfo {pages} {127}},\ \Eprint {https://arxiv.org/abs/1701.07436}
  {arXiv:1701.07436 [hep-th]} \BibitemShut {NoStop}%
%%CITATION = ARXIV:1701.07436;%%
\bibitem [{\citenamefont {Cao}\ \emph {et~al.}(2010)\citenamefont {Cao},
  \citenamefont {Elliott}, \citenamefont {Joseph}, \citenamefont {Wu},
  \citenamefont {Petricka}, \citenamefont {Schafer},\ and\ \citenamefont
  {Thomas}}]{Cao_2010}%
  \BibitemOpen
  \bibfield  {author} {\bibinfo {author} {\bibfnamefont {C.}~\bibnamefont
  {Cao}}, \bibinfo {author} {\bibfnamefont {E.}~\bibnamefont {Elliott}},
  \bibinfo {author} {\bibfnamefont {J.}~\bibnamefont {Joseph}}, \bibinfo
  {author} {\bibfnamefont {H.}~\bibnamefont {Wu}}, \bibinfo {author}
  {\bibfnamefont {J.}~\bibnamefont {Petricka}}, \bibinfo {author}
  {\bibfnamefont {T.}~\bibnamefont {Schafer}},\ and\ \bibinfo {author}
  {\bibfnamefont {J.~E.}\ \bibnamefont {Thomas}},\ }\bibfield  {title}
  {\bibinfo {title} {Universal quantum viscosity in a unitary fermi gas},\
  }\href {https://doi.org/10.1126/science.1195219} {\bibfield  {journal}
  {\bibinfo  {journal} {Science}\ }\textbf {\bibinfo {volume} {331}},\ \bibinfo
  {pages} {58} (\bibinfo {year} {2010})}\BibitemShut {NoStop}%
\bibitem [{\citenamefont {Shuryak}(2009)}]{Shuryak_2009}%
  \BibitemOpen
  \bibfield  {author} {\bibinfo {author} {\bibfnamefont {E.}~\bibnamefont
  {Shuryak}},\ }\bibfield  {title} {\bibinfo {title} {Physics of strongly
  coupled quark-gluon plasma},\ }\href
  {https://doi.org/10.1016/j.ppnp.2008.09.001} {\bibfield  {journal} {\bibinfo
  {journal} {Progress in Particle and Nuclear Physics}\ }\textbf {\bibinfo
  {volume} {62}},\ \bibinfo {pages} {48} (\bibinfo {year} {2009})}\BibitemShut
  {NoStop}%
\bibitem [{\citenamefont {Crossno}\ \emph {et~al.}(2016)\citenamefont
  {Crossno}, \citenamefont {Shi}, \citenamefont {Wang}, \citenamefont {Liu},
  \citenamefont {Harzheim}, \citenamefont {Lucas}, \citenamefont {Sachdev},
  \citenamefont {Kim}, \citenamefont {Taniguchi}, \citenamefont {Watanabe},\
  and\ \citenamefont {et~al.}}]{Crossno_2016}%
  \BibitemOpen
  \bibfield  {author} {\bibinfo {author} {\bibfnamefont {J.}~\bibnamefont
  {Crossno}}, \bibinfo {author} {\bibfnamefont {J.~K.}\ \bibnamefont {Shi}},
  \bibinfo {author} {\bibfnamefont {K.}~\bibnamefont {Wang}}, \bibinfo {author}
  {\bibfnamefont {X.}~\bibnamefont {Liu}}, \bibinfo {author} {\bibfnamefont
  {A.}~\bibnamefont {Harzheim}}, \bibinfo {author} {\bibfnamefont
  {A.}~\bibnamefont {Lucas}}, \bibinfo {author} {\bibfnamefont
  {S.}~\bibnamefont {Sachdev}}, \bibinfo {author} {\bibfnamefont
  {P.}~\bibnamefont {Kim}}, \bibinfo {author} {\bibfnamefont {T.}~\bibnamefont
  {Taniguchi}}, \bibinfo {author} {\bibfnamefont {K.}~\bibnamefont
  {Watanabe}},\ and\ \bibinfo {author} {\bibnamefont {et~al.}},\ }\bibfield
  {title} {\bibinfo {title} {{Observation of the Dirac fluid and the breakdown
  of the Wiedemann-Franz law in graphene}},\ }\href
  {https://doi.org/10.1126/science.aad0343} {\bibfield  {journal} {\bibinfo
  {journal} {Science}\ }\textbf {\bibinfo {volume} {351}},\ \bibinfo {pages}
  {1058} (\bibinfo {year} {2016})}\BibitemShut {NoStop}%
\bibitem [{\citenamefont {et~al.}(2016)}]{Bandurin_2016}%
  \BibitemOpen
  \bibfield  {author} {\bibinfo {author} {\bibfnamefont {D.~A.~B.}\
  \bibnamefont {et~al.}},\ }\bibfield  {title} {\bibinfo {title} {Negative
  local resistance caused by viscous electron backflow in graphene},\ }\href
  {https://doi.org/10.1126/science.aad0201} {\bibfield  {journal} {\bibinfo
  {journal} {Science}\ }\textbf {\bibinfo {volume} {351}},\ \bibinfo {pages}
  {1055} (\bibinfo {year} {2016})}\BibitemShut {NoStop}%
\bibitem [{\citenamefont {de~Jong}\ and\ \citenamefont
  {Molenkamp}(1995)}]{de_Jong_1995}%
  \BibitemOpen
  \bibfield  {author} {\bibinfo {author} {\bibfnamefont {M.~J.~M.}\
  \bibnamefont {de~Jong}}\ and\ \bibinfo {author} {\bibfnamefont {L.~W.}\
  \bibnamefont {Molenkamp}},\ }\bibfield  {title} {\bibinfo {title}
  {Hydrodynamic electron flow in high-mobility wires},\ }\href
  {https://doi.org/10.1103/physrevb.51.13389} {\bibfield  {journal} {\bibinfo
  {journal} {Physical Review B}\ }\textbf {\bibinfo {volume} {51}},\ \bibinfo
  {pages} {13389} (\bibinfo {year} {1995})}\BibitemShut {NoStop}%
\bibitem [{\citenamefont {et~al}(2017)}]{Krishna_Kumar_2017}%
  \BibitemOpen
  \bibfield  {author} {\bibinfo {author} {\bibfnamefont {R.~K.~K.}\
  \bibnamefont {et~al}},\ }\bibfield  {title} {\bibinfo {title} {Superballistic
  flow of viscous electron fluid through graphene constrictions},\ }\href
  {https://doi.org/10.1038/nphys4240} {\bibfield  {journal} {\bibinfo
  {journal} {Nature Physics}\ }\textbf {\bibinfo {volume} {13}},\ \bibinfo
  {pages} {1182} (\bibinfo {year} {2017})}\BibitemShut {NoStop}%
\bibitem [{\citenamefont {Chamon}(2005)}]{chamon2005quantum}%
  \BibitemOpen
  \bibfield  {author} {\bibinfo {author} {\bibfnamefont {C.}~\bibnamefont
  {Chamon}},\ }\bibfield  {title} {\bibinfo {title} {Quantum glassiness in
  strongly correlated clean systems: an example of topological
  overprotection},\ }\href@noop {} {\bibfield  {journal} {\bibinfo  {journal}
  {Physical review letters}\ }\textbf {\bibinfo {volume} {94}},\ \bibinfo
  {pages} {040402} (\bibinfo {year} {2005})}\BibitemShut {NoStop}%
\bibitem [{\citenamefont {Vijay}\ \emph {et~al.}(2015)\citenamefont {Vijay},
  \citenamefont {Haah},\ and\ \citenamefont {Fu}}]{vijay2015new}%
  \BibitemOpen
  \bibfield  {author} {\bibinfo {author} {\bibfnamefont {S.}~\bibnamefont
  {Vijay}}, \bibinfo {author} {\bibfnamefont {J.}~\bibnamefont {Haah}},\ and\
  \bibinfo {author} {\bibfnamefont {L.}~\bibnamefont {Fu}},\ }\bibfield
  {title} {\bibinfo {title} {A new kind of topological quantum order: A
  dimensional hierarchy of quasiparticles built from stationary excitations},\
  }\href@noop {} {\bibfield  {journal} {\bibinfo  {journal} {Physical Review
  B}\ }\textbf {\bibinfo {volume} {92}},\ \bibinfo {pages} {235136} (\bibinfo
  {year} {2015})}\BibitemShut {NoStop}%
\bibitem [{\citenamefont {Vijay}\ \emph {et~al.}(2016)\citenamefont {Vijay},
  \citenamefont {Haah},\ and\ \citenamefont {Fu}}]{vijay2016fracton}%
  \BibitemOpen
  \bibfield  {author} {\bibinfo {author} {\bibfnamefont {S.}~\bibnamefont
  {Vijay}}, \bibinfo {author} {\bibfnamefont {J.}~\bibnamefont {Haah}},\ and\
  \bibinfo {author} {\bibfnamefont {L.}~\bibnamefont {Fu}},\ }\bibfield
  {title} {\bibinfo {title} {Fracton topological order, generalized lattice
  gauge theory, and duality},\ }\href@noop {} {\bibfield  {journal} {\bibinfo
  {journal} {Physical Review B}\ }\textbf {\bibinfo {volume} {94}},\ \bibinfo
  {pages} {235157} (\bibinfo {year} {2016})}\BibitemShut {NoStop}%
\bibitem [{\citenamefont
  {Pretko}(2017{\natexlab{a}})}]{pretko2017subdimensional}%
  \BibitemOpen
  \bibfield  {author} {\bibinfo {author} {\bibfnamefont {M.}~\bibnamefont
  {Pretko}},\ }\bibfield  {title} {\bibinfo {title} {{Subdimensional particle
  structure of higher rank U(1) spin liquids}},\ }\href@noop {} {\bibfield
  {journal} {\bibinfo  {journal} {Physical Review B}\ }\textbf {\bibinfo
  {volume} {95}},\ \bibinfo {pages} {115139} (\bibinfo {year}
  {2017}{\natexlab{a}})}\BibitemShut {NoStop}%
\bibitem [{\citenamefont {Pretko}(2017{\natexlab{b}})}]{pretko2017emergent}%
  \BibitemOpen
  \bibfield  {author} {\bibinfo {author} {\bibfnamefont {M.}~\bibnamefont
  {Pretko}},\ }\bibfield  {title} {\bibinfo {title} {Emergent gravity of
  fractons: Mach's principle revisited},\ }\href@noop {} {\bibfield  {journal}
  {\bibinfo  {journal} {Physical Review D}\ }\textbf {\bibinfo {volume} {96}},\
  \bibinfo {pages} {024051} (\bibinfo {year} {2017}{\natexlab{b}})}\BibitemShut
  {NoStop}%
\bibitem [{\citenamefont {Pretko}(2017{\natexlab{c}})}]{pretko2017generalized}%
  \BibitemOpen
  \bibfield  {author} {\bibinfo {author} {\bibfnamefont {M.}~\bibnamefont
  {Pretko}},\ }\bibfield  {title} {\bibinfo {title} {Generalized
  electromagnetism of subdimensional particles: A spin liquid story},\
  }\href@noop {} {\bibfield  {journal} {\bibinfo  {journal} {Physical Review
  B}\ }\textbf {\bibinfo {volume} {96}},\ \bibinfo {pages} {035119} (\bibinfo
  {year} {2017}{\natexlab{c}})}\BibitemShut {NoStop}%
\bibitem [{\citenamefont {Slagle}\ and\ \citenamefont
  {Kim}(2017)}]{slagle2017fracton}%
  \BibitemOpen
  \bibfield  {author} {\bibinfo {author} {\bibfnamefont {K.}~\bibnamefont
  {Slagle}}\ and\ \bibinfo {author} {\bibfnamefont {Y.~B.}\ \bibnamefont
  {Kim}},\ }\bibfield  {title} {\bibinfo {title} {Fracton topological order
  from nearest-neighbor two-spin interactions and dualities},\ }\href@noop {}
  {\bibfield  {journal} {\bibinfo  {journal} {Physical Review B}\ }\textbf
  {\bibinfo {volume} {96}},\ \bibinfo {pages} {165106} (\bibinfo {year}
  {2017})}\BibitemShut {NoStop}%
\bibitem [{\citenamefont {Slagle}\ \emph {et~al.}(2018)\citenamefont {Slagle},
  \citenamefont {Prem},\ and\ \citenamefont {Pretko}}]{slagle2018symmetric}%
  \BibitemOpen
  \bibfield  {author} {\bibinfo {author} {\bibfnamefont {K.}~\bibnamefont
  {Slagle}}, \bibinfo {author} {\bibfnamefont {A.}~\bibnamefont {Prem}},\ and\
  \bibinfo {author} {\bibfnamefont {M.}~\bibnamefont {Pretko}},\ }\bibfield
  {title} {\bibinfo {title} {Symmetric tensor gauge theories on curved
  spaces},\ }\href@noop {} {\bibfield  {journal} {\bibinfo  {journal} {arXiv
  preprint arXiv:1807.00827}\ } (\bibinfo {year} {2018})}\BibitemShut {NoStop}%
\bibitem [{\citenamefont {Schmitz}\ \emph {et~al.}(2018)\citenamefont
  {Schmitz}, \citenamefont {Ma}, \citenamefont {Nandkishore},\ and\
  \citenamefont {Parameswaran}}]{schmitz2018recoverable}%
  \BibitemOpen
  \bibfield  {author} {\bibinfo {author} {\bibfnamefont {A.}~\bibnamefont
  {Schmitz}}, \bibinfo {author} {\bibfnamefont {H.}~\bibnamefont {Ma}},
  \bibinfo {author} {\bibfnamefont {R.~M.}\ \bibnamefont {Nandkishore}},\ and\
  \bibinfo {author} {\bibfnamefont {S.}~\bibnamefont {Parameswaran}},\
  }\bibfield  {title} {\bibinfo {title} {Recoverable information and emergent
  conservation laws in fracton stabilizer codes},\ }\href@noop {} {\bibfield
  {journal} {\bibinfo  {journal} {Physical Review B}\ }\textbf {\bibinfo
  {volume} {97}},\ \bibinfo {pages} {134426} (\bibinfo {year}
  {2018})}\BibitemShut {NoStop}%
\bibitem [{\citenamefont {Ma}\ \emph {et~al.}(2017)\citenamefont {Ma},
  \citenamefont {Lake}, \citenamefont {Chen},\ and\ \citenamefont
  {Hermele}}]{ma2017fracton}%
  \BibitemOpen
  \bibfield  {author} {\bibinfo {author} {\bibfnamefont {H.}~\bibnamefont
  {Ma}}, \bibinfo {author} {\bibfnamefont {E.}~\bibnamefont {Lake}}, \bibinfo
  {author} {\bibfnamefont {X.}~\bibnamefont {Chen}},\ and\ \bibinfo {author}
  {\bibfnamefont {M.}~\bibnamefont {Hermele}},\ }\bibfield  {title} {\bibinfo
  {title} {Fracton topological order via coupled layers},\ }\href@noop {}
  {\bibfield  {journal} {\bibinfo  {journal} {Physical Review B}\ }\textbf
  {\bibinfo {volume} {95}},\ \bibinfo {pages} {245126} (\bibinfo {year}
  {2017})}\BibitemShut {NoStop}%
\bibitem [{\citenamefont {Ma}\ \emph {et~al.}(2018)\citenamefont {Ma},
  \citenamefont {Schmitz}, \citenamefont {Parameswaran}, \citenamefont
  {Hermele},\ and\ \citenamefont {Nandkishore}}]{ma2018topological}%
  \BibitemOpen
  \bibfield  {author} {\bibinfo {author} {\bibfnamefont {H.}~\bibnamefont
  {Ma}}, \bibinfo {author} {\bibfnamefont {A.}~\bibnamefont {Schmitz}},
  \bibinfo {author} {\bibfnamefont {S.}~\bibnamefont {Parameswaran}}, \bibinfo
  {author} {\bibfnamefont {M.}~\bibnamefont {Hermele}},\ and\ \bibinfo {author}
  {\bibfnamefont {R.~M.}\ \bibnamefont {Nandkishore}},\ }\bibfield  {title}
  {\bibinfo {title} {Topological entanglement entropy of fracton stabilizer
  codes},\ }\href@noop {} {\bibfield  {journal} {\bibinfo  {journal} {Physical
  Review B}\ }\textbf {\bibinfo {volume} {97}},\ \bibinfo {pages} {125101}
  (\bibinfo {year} {2018})}\BibitemShut {NoStop}%
\bibitem [{\citenamefont {Shirley}\ \emph {et~al.}(2019)\citenamefont
  {Shirley}, \citenamefont {Slagle},\ and\ \citenamefont
  {Chen}}]{shirley2019foliated}%
  \BibitemOpen
  \bibfield  {author} {\bibinfo {author} {\bibfnamefont {W.}~\bibnamefont
  {Shirley}}, \bibinfo {author} {\bibfnamefont {K.}~\bibnamefont {Slagle}},\
  and\ \bibinfo {author} {\bibfnamefont {X.}~\bibnamefont {Chen}},\ }\bibfield
  {title} {\bibinfo {title} {Foliated fracton order from gauging subsystem
  symmetries},\ }\href@noop {} {\bibfield  {journal} {\bibinfo  {journal}
  {SciPost Physics}\ }\textbf {\bibinfo {volume} {6}},\ \bibinfo {pages} {041}
  (\bibinfo {year} {2019})}\BibitemShut {NoStop}%
\bibitem [{\citenamefont {Pretko}\ and\ \citenamefont
  {Radzihovsky}(2018)}]{Pretko2018}%
  \BibitemOpen
  \bibfield  {author} {\bibinfo {author} {\bibfnamefont {M.}~\bibnamefont
  {Pretko}}\ and\ \bibinfo {author} {\bibfnamefont {L.}~\bibnamefont
  {Radzihovsky}},\ }\bibfield  {title} {\bibinfo {title} {Fracton-elasticity
  duality},\ }\href {https://doi.org/10.1103/PhysRevLett.120.195301} {\bibfield
   {journal} {\bibinfo  {journal} {Phys. Rev. Lett.}\ }\textbf {\bibinfo
  {volume} {120}},\ \bibinfo {pages} {195301} (\bibinfo {year}
  {2018})}\BibitemShut {NoStop}%
\bibitem [{\citenamefont {Radzihovsky}\ and\ \citenamefont
  {Hermele}(2020)}]{radzihovsky2020fractons}%
  \BibitemOpen
  \bibfield  {author} {\bibinfo {author} {\bibfnamefont {L.}~\bibnamefont
  {Radzihovsky}}\ and\ \bibinfo {author} {\bibfnamefont {M.}~\bibnamefont
  {Hermele}},\ }\bibfield  {title} {\bibinfo {title} {Fractons from vector
  gauge theory},\ }\href@noop {} {\bibfield  {journal} {\bibinfo  {journal}
  {Physical Review Letters}\ }\textbf {\bibinfo {volume} {124}},\ \bibinfo
  {pages} {050402} (\bibinfo {year} {2020})}\BibitemShut {NoStop}%
\bibitem [{\citenamefont {Seiberg}\ and\ \citenamefont
  {Shao}(2020)}]{Seiberg_2020}%
  \BibitemOpen
  \bibfield  {author} {\bibinfo {author} {\bibfnamefont {N.}~\bibnamefont
  {Seiberg}}\ and\ \bibinfo {author} {\bibfnamefont {S.-H.}\ \bibnamefont
  {Shao}},\ }\bibfield  {title} {\bibinfo {title} {Exotic $u(1)$ symmetries,
  duality, and fractons in 3+1-dimensional quantum field theory},\ }\href
  {http://dx.doi.org/10.21468/SciPostPhys.9.4.046} {\bibfield  {journal}
  {\bibinfo  {journal} {SciPost Physics}\ }\textbf {\bibinfo {volume} {9}},\
  \bibinfo {pages} {046} (\bibinfo {year} {2020})}\BibitemShut {NoStop}%
\bibitem [{\citenamefont {Gorantla}\ \emph {et~al.}(2020)\citenamefont
  {Gorantla}, \citenamefont {Lam}, \citenamefont {Seiberg},\ and\ \citenamefont
  {Shao}}]{Gorantla_2020}%
  \BibitemOpen
  \bibfield  {author} {\bibinfo {author} {\bibfnamefont {P.}~\bibnamefont
  {Gorantla}}, \bibinfo {author} {\bibfnamefont {H.~T.}\ \bibnamefont {Lam}},
  \bibinfo {author} {\bibfnamefont {N.}~\bibnamefont {Seiberg}},\ and\ \bibinfo
  {author} {\bibfnamefont {S.-H.}\ \bibnamefont {Shao}},\ }\bibfield  {title}
  {\bibinfo {title} {More exotic field theories in 3+1 dimensions},\ }\href
  {http://dx.doi.org/10.21468/SciPostPhys.9.5.073} {\bibfield  {journal}
  {\bibinfo  {journal} {SciPost Physics}\ }\textbf {\bibinfo {volume} {9}},\
  \bibinfo {pages} {073} (\bibinfo {year} {2020})}\BibitemShut {NoStop}%
\bibitem [{\citenamefont {Rudelius}\ \emph {et~al.}(2020)\citenamefont
  {Rudelius}, \citenamefont {Seiberg},\ and\ \citenamefont
  {Shao}}]{rudelius2020fractons}%
  \BibitemOpen
  \bibfield  {author} {\bibinfo {author} {\bibfnamefont {T.}~\bibnamefont
  {Rudelius}}, \bibinfo {author} {\bibfnamefont {N.}~\bibnamefont {Seiberg}},\
  and\ \bibinfo {author} {\bibfnamefont {S.-H.}\ \bibnamefont {Shao}},\
  }\href@noop {} {\bibinfo {title} {Fractons with twisted boundary conditions
  and their symmetries}} (\bibinfo {year} {2020}),\ \Eprint
  {https://arxiv.org/abs/2012.11592} {arXiv:2012.11592 [cond-mat.str-el]}
  \BibitemShut {NoStop}%
\bibitem [{\citenamefont {Doshi}\ and\ \citenamefont {Gromov}(2021)}]{doshi}%
  \BibitemOpen
  \bibfield  {author} {\bibinfo {author} {\bibfnamefont {D.}~\bibnamefont
  {Doshi}}\ and\ \bibinfo {author} {\bibfnamefont {A.}~\bibnamefont {Gromov}},\
  }\bibfield  {title} {\bibinfo {title} {Vortices as fractons},\ }\href
  {https://doi.org/10.1038/s42005-021-00540-4} {\bibfield  {journal} {\bibinfo
  {journal} {Communications Physics}\ }\textbf {\bibinfo {volume} {4}},\
  \bibinfo {pages} {44} (\bibinfo {year} {2021})}\BibitemShut {NoStop}%
\bibitem [{\citenamefont {Pai}\ \emph {et~al.}(2019)\citenamefont {Pai},
  \citenamefont {Pretko},\ and\ \citenamefont
  {Nandkishore}}]{pai2019localization}%
  \BibitemOpen
  \bibfield  {author} {\bibinfo {author} {\bibfnamefont {S.}~\bibnamefont
  {Pai}}, \bibinfo {author} {\bibfnamefont {M.}~\bibnamefont {Pretko}},\ and\
  \bibinfo {author} {\bibfnamefont {R.~M.}\ \bibnamefont {Nandkishore}},\
  }\bibfield  {title} {\bibinfo {title} {Localization in fractonic random
  circuits},\ }\href@noop {} {\bibfield  {journal} {\bibinfo  {journal}
  {Physical Review X}\ }\textbf {\bibinfo {volume} {9}},\ \bibinfo {pages}
  {021003} (\bibinfo {year} {2019})}\BibitemShut {NoStop}%
\bibitem [{\citenamefont {Sala}\ \emph {et~al.}(2020)\citenamefont {Sala},
  \citenamefont {Rakovszky}, \citenamefont {Verresen}, \citenamefont {Knap},\
  and\ \citenamefont {Pollmann}}]{2020PollmannFragmentation}%
  \BibitemOpen
  \bibfield  {author} {\bibinfo {author} {\bibfnamefont {P.}~\bibnamefont
  {Sala}}, \bibinfo {author} {\bibfnamefont {T.}~\bibnamefont {Rakovszky}},
  \bibinfo {author} {\bibfnamefont {R.}~\bibnamefont {Verresen}}, \bibinfo
  {author} {\bibfnamefont {M.}~\bibnamefont {Knap}},\ and\ \bibinfo {author}
  {\bibfnamefont {F.}~\bibnamefont {Pollmann}},\ }\bibfield  {title} {\bibinfo
  {title} {{Ergodicity Breaking Arising from Hilbert Space Fragmentation in
  Dipole-Conserving Hamiltonians}},\ }\href
  {https://doi.org/10.1103/PhysRevX.10.011047} {\bibfield  {journal} {\bibinfo
  {journal} {Phys. Rev. X}\ }\textbf {\bibinfo {volume} {10}},\ \bibinfo
  {pages} {011047} (\bibinfo {year} {2020})}\BibitemShut {NoStop}%
\bibitem [{\citenamefont {Khemani}\ \emph {et~al.}(2020)\citenamefont
  {Khemani}, \citenamefont {Hermele},\ and\ \citenamefont
  {Nandkishore}}]{shattering}%
  \BibitemOpen
  \bibfield  {author} {\bibinfo {author} {\bibfnamefont {V.}~\bibnamefont
  {Khemani}}, \bibinfo {author} {\bibfnamefont {M.}~\bibnamefont {Hermele}},\
  and\ \bibinfo {author} {\bibfnamefont {R.}~\bibnamefont {Nandkishore}},\
  }\bibfield  {title} {\bibinfo {title} {Localization from hilbert space
  shattering: From theory to physical realizations},\ }\href
  {https://doi.org/10.1103/PhysRevB.101.174204} {\bibfield  {journal} {\bibinfo
   {journal} {Phys. Rev. B}\ }\textbf {\bibinfo {volume} {101}},\ \bibinfo
  {pages} {174204} (\bibinfo {year} {2020})}\BibitemShut {NoStop}%
\bibitem [{\citenamefont {Gromov}\ \emph {et~al.}(2020)\citenamefont {Gromov},
  \citenamefont {Lucas},\ and\ \citenamefont {Nandkishore}}]{fractonhydro}%
  \BibitemOpen
  \bibfield  {author} {\bibinfo {author} {\bibfnamefont {A.}~\bibnamefont
  {Gromov}}, \bibinfo {author} {\bibfnamefont {A.}~\bibnamefont {Lucas}},\ and\
  \bibinfo {author} {\bibfnamefont {R.~M.}\ \bibnamefont {Nandkishore}},\
  }\bibfield  {title} {\bibinfo {title} {Fracton hydrodynamics},\ }\href
  {https://doi.org/10.1103/PhysRevResearch.2.033124} {\bibfield  {journal}
  {\bibinfo  {journal} {Phys. Rev. Research}\ }\textbf {\bibinfo {volume}
  {2}},\ \bibinfo {pages} {033124} (\bibinfo {year} {2020})}\BibitemShut
  {NoStop}%
\bibitem [{\citenamefont {Feldmeier}\ \emph {et~al.}(2020)\citenamefont
  {Feldmeier}, \citenamefont {Sala}, \citenamefont {De~Tomasi}, \citenamefont
  {Pollmann},\ and\ \citenamefont {Knap}}]{knap2020}%
  \BibitemOpen
  \bibfield  {author} {\bibinfo {author} {\bibfnamefont {J.}~\bibnamefont
  {Feldmeier}}, \bibinfo {author} {\bibfnamefont {P.}~\bibnamefont {Sala}},
  \bibinfo {author} {\bibfnamefont {G.}~\bibnamefont {De~Tomasi}}, \bibinfo
  {author} {\bibfnamefont {F.}~\bibnamefont {Pollmann}},\ and\ \bibinfo
  {author} {\bibfnamefont {M.}~\bibnamefont {Knap}},\ }\bibfield  {title}
  {\bibinfo {title} {Anomalous diffusion in dipole- and
  higher-moment-conserving systems},\ }\href
  {https://doi.org/10.1103/PhysRevLett.125.245303} {\bibfield  {journal}
  {\bibinfo  {journal} {Phys. Rev. Lett.}\ }\textbf {\bibinfo {volume} {125}},\
  \bibinfo {pages} {245303} (\bibinfo {year} {2020})}\BibitemShut {NoStop}%
\bibitem [{\citenamefont {Zhang}(2020{\natexlab{a}})}]{zhang2020}%
  \BibitemOpen
  \bibfield  {author} {\bibinfo {author} {\bibfnamefont {P.}~\bibnamefont
  {Zhang}},\ }\bibfield  {title} {\bibinfo {title} {Subdiffusion in strongly
  tilted lattice systems},\ }\href
  {https://doi.org/10.1103/PhysRevResearch.2.033129} {\bibfield  {journal}
  {\bibinfo  {journal} {Phys. Rev. Research}\ }\textbf {\bibinfo {volume}
  {2}},\ \bibinfo {pages} {033129} (\bibinfo {year}
  {2020}{\natexlab{a}})}\BibitemShut {NoStop}%
\bibitem [{\citenamefont {Morningstar}\ \emph {et~al.}(2020)\citenamefont
  {Morningstar}, \citenamefont {Khemani},\ and\ \citenamefont
  {Huse}}]{morningstar}%
  \BibitemOpen
  \bibfield  {author} {\bibinfo {author} {\bibfnamefont {A.}~\bibnamefont
  {Morningstar}}, \bibinfo {author} {\bibfnamefont {V.}~\bibnamefont
  {Khemani}},\ and\ \bibinfo {author} {\bibfnamefont {D.~A.}\ \bibnamefont
  {Huse}},\ }\bibfield  {title} {\bibinfo {title} {Kinetically constrained
  freezing transition in a dipole-conserving system},\ }\href
  {https://doi.org/10.1103/PhysRevB.101.214205} {\bibfield  {journal} {\bibinfo
   {journal} {Phys. Rev. B}\ }\textbf {\bibinfo {volume} {101}},\ \bibinfo
  {pages} {214205} (\bibinfo {year} {2020})}\BibitemShut {NoStop}%
\bibitem [{\citenamefont {Iaconis}\ \emph {et~al.}(2021)\citenamefont
  {Iaconis}, \citenamefont {Lucas},\ and\ \citenamefont
  {Nandkishore}}]{iaconis2021}%
  \BibitemOpen
  \bibfield  {author} {\bibinfo {author} {\bibfnamefont {J.}~\bibnamefont
  {Iaconis}}, \bibinfo {author} {\bibfnamefont {A.}~\bibnamefont {Lucas}},\
  and\ \bibinfo {author} {\bibfnamefont {R.}~\bibnamefont {Nandkishore}},\
  }\bibfield  {title} {\bibinfo {title} {Multipole conservation laws and
  subdiffusion in any dimension},\ }\href
  {https://doi.org/10.1103/PhysRevE.103.022142} {\bibfield  {journal} {\bibinfo
   {journal} {Phys. Rev. E}\ }\textbf {\bibinfo {volume} {103}},\ \bibinfo
  {pages} {022142} (\bibinfo {year} {2021})}\BibitemShut {NoStop}%
\bibitem [{\citenamefont {Ganesan}\ and\ \citenamefont
  {Lucas}(2020)}]{Ganesan:2020wvm}%
  \BibitemOpen
  \bibfield  {author} {\bibinfo {author} {\bibfnamefont {K.}~\bibnamefont
  {Ganesan}}\ and\ \bibinfo {author} {\bibfnamefont {A.}~\bibnamefont
  {Lucas}},\ }\bibfield  {title} {\bibinfo {title} {{Holographic
  subdiffusion}},\ }\href {https://doi.org/10.1007/JHEP12(2020)149} {\bibfield
  {journal} {\bibinfo  {journal} {JHEP}\ }\textbf {\bibinfo {volume} {12}},\
  \bibinfo {pages} {149}},\ \Eprint {https://arxiv.org/abs/2008.09638}
  {arXiv:2008.09638 [hep-th]} \BibitemShut {NoStop}%
\bibitem [{\citenamefont {Guardado-Sanchez}\ \emph {et~al.}(2020)\citenamefont
  {Guardado-Sanchez}, \citenamefont {Morningstar}, \citenamefont {Spar},
  \citenamefont {Brown}, \citenamefont {Huse},\ and\ \citenamefont
  {Bakr}}]{Guardado_Sanchez_2020}%
  \BibitemOpen
  \bibfield  {author} {\bibinfo {author} {\bibfnamefont {E.}~\bibnamefont
  {Guardado-Sanchez}}, \bibinfo {author} {\bibfnamefont {A.}~\bibnamefont
  {Morningstar}}, \bibinfo {author} {\bibfnamefont {B.~M.}\ \bibnamefont
  {Spar}}, \bibinfo {author} {\bibfnamefont {P.~T.}\ \bibnamefont {Brown}},
  \bibinfo {author} {\bibfnamefont {D.~A.}\ \bibnamefont {Huse}},\ and\
  \bibinfo {author} {\bibfnamefont {W.~S.}\ \bibnamefont {Bakr}},\ }\bibfield
  {title} {\bibinfo {title} {{Subdiffusion and Heat Transport in a Tilted
  Two-Dimensional Fermi-Hubbard System}},\ }\href
  {https://doi.org/10.1103/physrevx.10.011042} {\bibfield  {journal} {\bibinfo
  {journal} {Physical Review X}\ }\textbf {\bibinfo {volume} {10}},\ \bibinfo
  {pages} {011042} (\bibinfo {year} {2020})}\BibitemShut {NoStop}%
\bibitem [{\citenamefont {Kardar}\ \emph {et~al.}(1986)\citenamefont {Kardar},
  \citenamefont {Parisi},\ and\ \citenamefont {Zhang}}]{kpz}%
  \BibitemOpen
  \bibfield  {author} {\bibinfo {author} {\bibfnamefont {M.}~\bibnamefont
  {Kardar}}, \bibinfo {author} {\bibfnamefont {G.}~\bibnamefont {Parisi}},\
  and\ \bibinfo {author} {\bibfnamefont {Y.-C.}\ \bibnamefont {Zhang}},\
  }\bibfield  {title} {\bibinfo {title} {Dynamic scaling of growing
  interfaces},\ }\href {https://doi.org/10.1103/PhysRevLett.56.889} {\bibfield
  {journal} {\bibinfo  {journal} {Phys. Rev. Lett.}\ }\textbf {\bibinfo
  {volume} {56}},\ \bibinfo {pages} {889} (\bibinfo {year} {1986})}\BibitemShut
  {NoStop}%
\bibitem [{\citenamefont {Delacr\'etaz}\ and\ \citenamefont
  {Glorioso}(2020)}]{gloriosoprl}%
  \BibitemOpen
  \bibfield  {author} {\bibinfo {author} {\bibfnamefont {L.~V.}\ \bibnamefont
  {Delacr\'etaz}}\ and\ \bibinfo {author} {\bibfnamefont {P.}~\bibnamefont
  {Glorioso}},\ }\bibfield  {title} {\bibinfo {title} {Breakdown of diffusion
  on chiral edges},\ }\href {https://doi.org/10.1103/PhysRevLett.124.236802}
  {\bibfield  {journal} {\bibinfo  {journal} {Phys. Rev. Lett.}\ }\textbf
  {\bibinfo {volume} {124}},\ \bibinfo {pages} {236802} (\bibinfo {year}
  {2020})}\BibitemShut {NoStop}%
\bibitem [{\citenamefont {Spohn}(2014)}]{Spohn_2014}%
  \BibitemOpen
  \bibfield  {author} {\bibinfo {author} {\bibfnamefont {H.}~\bibnamefont
  {Spohn}},\ }\bibfield  {title} {\bibinfo {title} {Nonlinear fluctuating
  hydrodynamics for anharmonic chains},\ }\href
  {https://doi.org/10.1007/s10955-014-0933-y} {\bibfield  {journal} {\bibinfo
  {journal} {Journal of Statistical Physics}\ }\textbf {\bibinfo {volume}
  {154}},\ \bibinfo {pages} {1191–1227} (\bibinfo {year} {2014})}\BibitemShut
  {NoStop}%
\bibitem [{\citenamefont {{Toner}}\ \emph {et~al.}(2005)\citenamefont
  {{Toner}}, \citenamefont {{Tu}},\ and\ \citenamefont
  {{Ramaswamy}}}]{flocking}%
  \BibitemOpen
  \bibfield  {author} {\bibinfo {author} {\bibfnamefont {J.}~\bibnamefont
  {{Toner}}}, \bibinfo {author} {\bibfnamefont {Y.}~\bibnamefont {{Tu}}},\ and\
  \bibinfo {author} {\bibfnamefont {S.}~\bibnamefont {{Ramaswamy}}},\
  }\bibfield  {title} {\bibinfo {title} {{Hydrodynamics and phases of
  flocks}},\ }\href {https://doi.org/10.1016/j.aop.2005.04.011} {\bibfield
  {journal} {\bibinfo  {journal} {Annals of Physics}\ }\textbf {\bibinfo
  {volume} {318}},\ \bibinfo {pages} {170} (\bibinfo {year}
  {2005})}\BibitemShut {NoStop}%
\bibitem [{\citenamefont {Hanai}\ and\ \citenamefont
  {Littlewood}(2020)}]{PhysRevResearch.2.033018}%
  \BibitemOpen
  \bibfield  {author} {\bibinfo {author} {\bibfnamefont {R.}~\bibnamefont
  {Hanai}}\ and\ \bibinfo {author} {\bibfnamefont {P.~B.}\ \bibnamefont
  {Littlewood}},\ }\bibfield  {title} {\bibinfo {title} {Critical fluctuations
  at a many-body exceptional point},\ }\href
  {https://doi.org/10.1103/PhysRevResearch.2.033018} {\bibfield  {journal}
  {\bibinfo  {journal} {Phys. Rev. Research}\ }\textbf {\bibinfo {volume}
  {2}},\ \bibinfo {pages} {033018} (\bibinfo {year} {2020})}\BibitemShut
  {NoStop}%
\bibitem [{\citenamefont {Mazenko}\ \emph {et~al.}(1983)\citenamefont
  {Mazenko}, \citenamefont {Ramaswamy},\ and\ \citenamefont {Toner}}]{toner}%
  \BibitemOpen
  \bibfield  {author} {\bibinfo {author} {\bibfnamefont {G.~F.}\ \bibnamefont
  {Mazenko}}, \bibinfo {author} {\bibfnamefont {S.}~\bibnamefont {Ramaswamy}},\
  and\ \bibinfo {author} {\bibfnamefont {J.}~\bibnamefont {Toner}},\ }\bibfield
   {title} {\bibinfo {title} {Breakdown of conventional hydrodynamics for
  smectic-$a$, hexatic-$b$, and cholesteric liquid crystals},\ }\href
  {https://doi.org/10.1103/PhysRevA.28.1618} {\bibfield  {journal} {\bibinfo
  {journal} {Phys. Rev. A}\ }\textbf {\bibinfo {volume} {28}},\ \bibinfo
  {pages} {1618} (\bibinfo {year} {1983})}\BibitemShut {NoStop}%
\bibitem [{\citenamefont {Simon}(1998)}]{simon}%
  \BibitemOpen
  \bibfield  {author} {\bibinfo {author} {\bibfnamefont {S.~H.}\ \bibnamefont
  {Simon}},\ }\bibfield  {title} {\bibinfo {title} {{The Chern-Simons Fermi
  liquid description of fractional quantum Hall states}},\ }in\ \href
  {https://doi.org/10.1142/9789812815989_0002} {\emph {\bibinfo {booktitle}
  {Composite Fermions}}},\ \bibinfo {editor} {edited by\ \bibinfo {editor}
  {\bibfnamefont {O.}~\bibnamefont {Heinonen}}}\ (\bibinfo  {publisher} {World
  Scientific},\ \bibinfo {year} {1998})\ p.\ \bibinfo {pages}
  {91–194}\BibitemShut {NoStop}%
\bibitem [{\citenamefont {Fliss}(2021)}]{fliss2021entanglement}%
  \BibitemOpen
  \bibfield  {author} {\bibinfo {author} {\bibfnamefont {J.~R.}\ \bibnamefont
  {Fliss}},\ }\href@noop {} {\bibinfo {title} {Entanglement in the quantum hall
  fluid of dipoles}} (\bibinfo {year} {2021}),\ \Eprint
  {https://arxiv.org/abs/2105.07448} {arXiv:2105.07448 [cond-mat.str-el]}
  \BibitemShut {NoStop}%
\bibitem [{\citenamefont {Gromov}(2019)}]{gromov2018towards}%
  \BibitemOpen
  \bibfield  {author} {\bibinfo {author} {\bibfnamefont {A.}~\bibnamefont
  {Gromov}},\ }\bibfield  {title} {\bibinfo {title} {Towards classification of
  fracton phases: the multipole algebra},\ }\href@noop {} {\bibfield  {journal}
  {\bibinfo  {journal} {Physical Review X}\ }\textbf {\bibinfo {volume} {9}},\
  \bibinfo {pages} {031035} (\bibinfo {year} {2019})}\BibitemShut {NoStop}%
\bibitem [{\citenamefont {Hohenberg}\ and\ \citenamefont
  {Halperin}(1977)}]{hohenberg}%
  \BibitemOpen
  \bibfield  {author} {\bibinfo {author} {\bibfnamefont {P.~C.}\ \bibnamefont
  {Hohenberg}}\ and\ \bibinfo {author} {\bibfnamefont {B.~I.}\ \bibnamefont
  {Halperin}},\ }\bibfield  {title} {\bibinfo {title} {Theory of dynamic
  critical phenomena},\ }\href {https://doi.org/10.1103/RevModPhys.49.435}
  {\bibfield  {journal} {\bibinfo  {journal} {Rev. Mod. Phys.}\ }\textbf
  {\bibinfo {volume} {49}},\ \bibinfo {pages} {435} (\bibinfo {year}
  {1977})}\BibitemShut {NoStop}%
\bibitem [{\citenamefont {Glorioso}\ \emph {et~al.}(2021)\citenamefont
  {Glorioso}, \citenamefont {Delacrétaz}, \citenamefont {Chen}, \citenamefont
  {Nandkishore},\ and\ \citenamefont {Lucas}}]{Glorioso_2021}%
  \BibitemOpen
  \bibfield  {author} {\bibinfo {author} {\bibfnamefont {P.}~\bibnamefont
  {Glorioso}}, \bibinfo {author} {\bibfnamefont {L.}~\bibnamefont
  {Delacrétaz}}, \bibinfo {author} {\bibfnamefont {X.}~\bibnamefont {Chen}},
  \bibinfo {author} {\bibfnamefont {R.}~\bibnamefont {Nandkishore}},\ and\
  \bibinfo {author} {\bibfnamefont {A.}~\bibnamefont {Lucas}},\ }\bibfield
  {title} {\bibinfo {title} {Hydrodynamics in lattice models with continuous
  non-abelian symmetries},\ }\href
  {http://dx.doi.org/10.21468/SciPostPhys.10.1.015} {\bibfield  {journal}
  {\bibinfo  {journal} {SciPost Physics}\ }\textbf {\bibinfo {volume} {10}},\
  \bibinfo {pages} {015} (\bibinfo {year} {2021})}\BibitemShut {NoStop}%
\bibitem [{\citenamefont {et~al.}()}]{future}%
  \BibitemOpen
  \bibfield  {author} {\bibinfo {author} {\bibfnamefont {P.~G.}\ \bibnamefont
  {et~al.}},\ }\href@noop {} {\bibinfo {title} {to appear}}\BibitemShut
  {NoStop}%
\bibitem [{\citenamefont {Forster}(1975)}]{forster}%
  \BibitemOpen
  \bibfield  {author} {\bibinfo {author} {\bibfnamefont {D.}~\bibnamefont
  {Forster}},\ }\href@noop {} {\emph {\bibinfo {title} {Hydrodynamic
  Fluctuations: Broken Symmetry and Correlation Functions}}}\ (\bibinfo
  {publisher} {Addison-Wesley},\ \bibinfo {year} {1975})\BibitemShut {NoStop}%
\bibitem [{\citenamefont {Zhang}(2020{\natexlab{b}})}]{zhang2020universal}%
  \BibitemOpen
  \bibfield  {author} {\bibinfo {author} {\bibfnamefont {P.}~\bibnamefont
  {Zhang}},\ }\href@noop {} {\bibinfo {title} {Universal subdiffusion in
  strongly tilted many-body systems}} (\bibinfo {year} {2020}{\natexlab{b}}),\
  \Eprint {https://arxiv.org/abs/2004.08695} {arXiv:2004.08695
  [cond-mat.quant-gas]} \BibitemShut {NoStop}%
\bibitem [{\citenamefont {Lepri}\ \emph {et~al.}(1997)\citenamefont {Lepri},
  \citenamefont {Livi},\ and\ \citenamefont {Politi}}]{PhysRevLett.78.1896}%
  \BibitemOpen
  \bibfield  {author} {\bibinfo {author} {\bibfnamefont {S.}~\bibnamefont
  {Lepri}}, \bibinfo {author} {\bibfnamefont {R.}~\bibnamefont {Livi}},\ and\
  \bibinfo {author} {\bibfnamefont {A.}~\bibnamefont {Politi}},\ }\bibfield
  {title} {\bibinfo {title} {Heat conduction in chains of nonlinear
  oscillators},\ }\href {https://doi.org/10.1103/PhysRevLett.78.1896}
  {\bibfield  {journal} {\bibinfo  {journal} {Phys. Rev. Lett.}\ }\textbf
  {\bibinfo {volume} {78}},\ \bibinfo {pages} {1896} (\bibinfo {year}
  {1997})}\BibitemShut {NoStop}%
\bibitem [{\citenamefont {Lee-Dadswell}\ \emph {et~al.}(2010)\citenamefont
  {Lee-Dadswell}, \citenamefont {Turner}, \citenamefont {Ettinger},\ and\
  \citenamefont {Moy}}]{PhysRevE.82.061118}%
  \BibitemOpen
  \bibfield  {author} {\bibinfo {author} {\bibfnamefont {G.~R.}\ \bibnamefont
  {Lee-Dadswell}}, \bibinfo {author} {\bibfnamefont {E.}~\bibnamefont
  {Turner}}, \bibinfo {author} {\bibfnamefont {J.}~\bibnamefont {Ettinger}},\
  and\ \bibinfo {author} {\bibfnamefont {M.}~\bibnamefont {Moy}},\ }\bibfield
  {title} {\bibinfo {title} {Momentum conserving one-dimensional system with a
  finite thermal conductivity},\ }\href
  {https://doi.org/10.1103/PhysRevE.82.061118} {\bibfield  {journal} {\bibinfo
  {journal} {Phys. Rev. E}\ }\textbf {\bibinfo {volume} {82}},\ \bibinfo
  {pages} {061118} (\bibinfo {year} {2010})}\BibitemShut {NoStop}%
\bibitem [{\citenamefont {Zhong}\ \emph {et~al.}(2012)\citenamefont {Zhong},
  \citenamefont {Zhang}, \citenamefont {Wang},\ and\ \citenamefont
  {Zhao}}]{PhysRevE.85.060102}%
  \BibitemOpen
  \bibfield  {author} {\bibinfo {author} {\bibfnamefont {Y.}~\bibnamefont
  {Zhong}}, \bibinfo {author} {\bibfnamefont {Y.}~\bibnamefont {Zhang}},
  \bibinfo {author} {\bibfnamefont {J.}~\bibnamefont {Wang}},\ and\ \bibinfo
  {author} {\bibfnamefont {H.}~\bibnamefont {Zhao}},\ }\bibfield  {title}
  {\bibinfo {title} {Normal heat conduction in one-dimensional momentum
  conserving lattices with asymmetric interactions},\ }\href
  {https://doi.org/10.1103/PhysRevE.85.060102} {\bibfield  {journal} {\bibinfo
  {journal} {Phys. Rev. E}\ }\textbf {\bibinfo {volume} {85}},\ \bibinfo
  {pages} {060102} (\bibinfo {year} {2012})}\BibitemShut {NoStop}%
\bibitem [{\citenamefont {Das}\ \emph {et~al.}(2014)\citenamefont {Das},
  \citenamefont {Dhar}, \citenamefont {Saito}, \citenamefont {Mendl},\ and\
  \citenamefont {Spohn}}]{PhysRevE.90.012124}%
  \BibitemOpen
  \bibfield  {author} {\bibinfo {author} {\bibfnamefont {S.~G.}\ \bibnamefont
  {Das}}, \bibinfo {author} {\bibfnamefont {A.}~\bibnamefont {Dhar}}, \bibinfo
  {author} {\bibfnamefont {K.}~\bibnamefont {Saito}}, \bibinfo {author}
  {\bibfnamefont {C.~B.}\ \bibnamefont {Mendl}},\ and\ \bibinfo {author}
  {\bibfnamefont {H.}~\bibnamefont {Spohn}},\ }\bibfield  {title} {\bibinfo
  {title} {Numerical test of hydrodynamic fluctuation theory in the
  fermi-pasta-ulam chain},\ }\href {https://doi.org/10.1103/PhysRevE.90.012124}
  {\bibfield  {journal} {\bibinfo  {journal} {Phys. Rev. E}\ }\textbf {\bibinfo
  {volume} {90}},\ \bibinfo {pages} {012124} (\bibinfo {year}
  {2014})}\BibitemShut {NoStop}%
\bibitem [{\citenamefont {Iaconis}\ \emph {et~al.}(2019)\citenamefont
  {Iaconis}, \citenamefont {Vijay},\ and\ \citenamefont
  {Nandkishore}}]{IaconisVijayNandkishore}%
  \BibitemOpen
  \bibfield  {author} {\bibinfo {author} {\bibfnamefont {J.}~\bibnamefont
  {Iaconis}}, \bibinfo {author} {\bibfnamefont {S.}~\bibnamefont {Vijay}},\
  and\ \bibinfo {author} {\bibfnamefont {R.}~\bibnamefont {Nandkishore}},\
  }\bibfield  {title} {\bibinfo {title} {Anomalous subdiffusion from subsystem
  symmetries},\ }\href {https://doi.org/10.1103/PhysRevB.100.214301} {\bibfield
   {journal} {\bibinfo  {journal} {Phys. Rev. B}\ }\textbf {\bibinfo {volume}
  {100}},\ \bibinfo {pages} {214301} (\bibinfo {year} {2019})}\BibitemShut
  {NoStop}%
\bibitem [{\citenamefont {Feldmeier}\ \emph {et~al.}(2021)\citenamefont
  {Feldmeier}, \citenamefont {Pollmann},\ and\ \citenamefont
  {Knap}}]{knap2021}%
  \BibitemOpen
  \bibfield  {author} {\bibinfo {author} {\bibfnamefont {J.}~\bibnamefont
  {Feldmeier}}, \bibinfo {author} {\bibfnamefont {F.}~\bibnamefont
  {Pollmann}},\ and\ \bibinfo {author} {\bibfnamefont {M.}~\bibnamefont
  {Knap}},\ }\bibfield  {title} {\bibinfo {title} {Emergent fracton dynamics in
  a nonplanar dimer model},\ }\href
  {https://doi.org/10.1103/PhysRevB.103.094303} {\bibfield  {journal} {\bibinfo
   {journal} {Phys. Rev. B}\ }\textbf {\bibinfo {volume} {103}},\ \bibinfo
  {pages} {094303} (\bibinfo {year} {2021})}\BibitemShut {NoStop}%
\bibitem [{\citenamefont {Grosvenor}\ \emph {et~al.}(2021)\citenamefont
  {Grosvenor}, \citenamefont {Hoyos}, \citenamefont {Pena-Benitez},\ and\
  \citenamefont {Surowka}}]{grosvenor2021hydrodynamics}%
  \BibitemOpen
  \bibfield  {author} {\bibinfo {author} {\bibfnamefont {K.~T.}\ \bibnamefont
  {Grosvenor}}, \bibinfo {author} {\bibfnamefont {C.}~\bibnamefont {Hoyos}},
  \bibinfo {author} {\bibfnamefont {F.}~\bibnamefont {Pena-Benitez}},\ and\
  \bibinfo {author} {\bibfnamefont {P.}~\bibnamefont {Surowka}},\ }\href@noop
  {} {\bibinfo {title} {Hydrodynamics of ideal fracton fluids}} (\bibinfo
  {year} {2021}),\ \Eprint {https://arxiv.org/abs/2105.01084} {arXiv:2105.01084
  [cond-mat.str-el]} \BibitemShut {NoStop}%
\end{thebibliography}%

\end{document}